\documentclass[12pt]{article}
\usepackage{epsfig,amsmath,amssymb}

\def\ra{\rangle}
\def\la{\langle}

\def\lsim{\raise0.3ex\hbox{$\;<$\kern-0.75em\raise-1.1ex\hbox{$\sim\;$}}}
\def\gsim{\raise0.3ex\hbox{$\;>$\kern-0.75em\raise-1.1ex\hbox{$\sim\;$}}}

\def\relic{\Omega_\chi}

\begin{document}

\vspace{-1truecm}

\rightline{CPHT--RR076.0908}
\rightline{LPT--Orsay 08/78}
%\rightline{hep-th/yymmnnn}
%\rightline{September 29th 2008}

\vspace{0.cm}

\begin{center}

{\Large {\bf Gauge vs. Gravity mediation in models with 
anomalous U(1)'s}}
\vspace{1 cm}\\

{\large E. Dudas$^{1,2}$, Y. Mambrini$^2$, S. Pokorski$^{3}$, A. Romagnoni$^{2,1}$, M. Trapletti$^{2,1}$
}
\vspace{1cm}\\

$^1$
CPhT, Ecole Polytechnique, CNRS, 91128 Palaiseau Cedex, France
\vspace{0.3cm}\\

$^2$
Laboratoire de Physique Th\'eorique,
Univ. Paris-Sud and CNRS, F-91405 Orsay, France
\vspace{0.3cm}\\

$^3$
Institute of Theoretical Physics, Warsaw University, Hoza 69, 00-681 Warsaw,
Poland
\vspace{0.3cm}\\

\end{center}

\vspace{1cm}

\abstract{\noindent
In an attempt to implement gauge mediation in  string theory, we study
string effective supergravity models of supersymmetry breaking, containing anomalous
gauge factors. We discuss subtleties related to gauge invariance and the stabilization
of the Green-Schwarz moduli, which set non-trivial constraints on the transmission of
supersymmetry breaking to MSSM via gauge interactions.
Given those constraints, it is difficult to obtain the dominance of gauge mediation over gravity
mediation. Furthermore,  generically the gauge contributions to soft
terms contain additional non-standard terms coming from D-term contributions.
Motivated by this, we study the phenomenology of recently proposed hybrid
models, where gravity and gauge mediations compete at the GUT scale, and show that such a
scenario can respect WMAP constraints and would be easily testable at LHC.
}

\newpage

\vspace{3cm}

\newpage

\tableofcontents

\vspace{3cm}

\pagestyle{plain}

\section{Introduction}

Recently there was some renewed interest in implementing gauge
mediation \cite{gr} in explicit supergravity framework and in string
theory \cite{dfks,aks,dmpr,cw,hmssv,hv}.
In models without anomalous $U(1)$ symmetries, gauge mediation contribution
to the soft supersymmetry breaking terms naturally dominates  over
gravity mediation contributions \cite{hybrid,lpt}.  This happens, for
example, adding messengers in models of F-term uplift \cite{lpt}.  
However, it is difficult to to find string origin of such models.
In string constructions, there are generically present anomalous $U(1)$
symmetries  and the messenger fields, which are typically
bifundamentals between SM gauge factors and the $U(1)$'s, are charged under them, as
well as the Green-Schwarz moduli responsible for anomaly  cancellation.  It is therefore important to study
if the dominance of gauge mediation can be obtained in models where $U(1)$ symmetry plays important role
in the supersymmetry breaking sector. Several explicit string models attempting to implement gauge mediation
in the presence of anomalous $U(1)$ have been recently constructed
\cite{dfks,aks,dmpr,cw,hmssv,hv} and it is
generally believed that this framework  can naturally provide a viable gauge mediation transmission of
supersymmetry breaking. However, as pointed out recently in a particular example \cite{dmpr}, there are
some subtleties to take into account if D-term contributions are present in this case.

The purpose of the present paper is to enter
more into the structure of these class of models and the various
constraints arising from:
\begin{itemize}
\item[i)] gauge invariance via the  Green-Schwarz mechanism;
\item[ii)] stabilization of the Green-Schwarz moduli fields;
\item[iii)] avoiding a Fayet-Iliopoulos term for the hypercharge and the existence
of non-standard gauge contributions to the MSSM soft terms, which naturally arise in the presence of
D-terms for $U(1)$'s under which messengers are charged, with important phenomenological consequences;
\end{itemize}

We show that the correct implementation of these constraints implies
that it is hard to parametrically strongly suppress gravity
contributions compared to the gauge ones, though some dominance of gauge mediation,
easing the FCNC problems of gravity, is possible in some regions of
the parameter space. Compared to uplifting models without $U(1)$'s,
the difference is that gauge invariance strongly correlates
supersymmetry breaking to the sector of moduli stabilization through
the gauge scalar potential. This translates into the fact that the
moduli-dependent Fayet-Iliopoulos term cannot be parametrically
much smaller than the Planck mass. Thus, the vev of the field $S$ breaking
SUSY is large\footnote{Most of the expressions are presented in Planck units,
namely, we set $M_{P} = 1$. However, at some instances, we keep
$M_{P}$ explicitly to make the discussions clearer.} $ 10^{-1} \le S \le  10^{-3}$ and as a result there is a
phenomenological lower limit on the gravitino mass $m_{3/2} \gsim 50-100$ GeV . In addition, there are
additional non-standard gauge mediation contributions. For example, constraint
iii) is hard to satisfy in some models with non-zero D-terms
\cite{aks,cw}, whereas in models with zero D-terms \cite{hv}
implementation of i) and ii) generically creates new supersymmetric
vacua.

In fact, once the above constraints are carefully taken into account, one is led to a class of ``hybrid'' models,
where both gravity mediation and gauge mediation  are equally important.
In the second part of this paper we therefore pursue in some detail a generalization of the model introduced in
 \cite{dmpr} where the above constraints are taken into account.

High energy physics approaches a new era with the
LHC. It will be possible to test a much larger part of the parameter
space of SUSY models, exclude some of them, or confirm the presence of supersymmetric
particles. Through the measurement of physical parameters (masses, couplings,
branching ratios), we will be able to extract fundamental informations
on the SUSY breaking mechanism: gauge, gravity, anomalous $U(1)$ mediation \cite{bd}, the anomaly mediation
\cite{anomaly}, gaugino mediation \cite{gaugino}, mirage mediation
\cite{Choi:2004sx}, etc. 

Here we point out that in the presence of anomalous $U(1)$ and  messenger fields,
charged under $U(1)$ and under the SM gauge groups, one naturally obtains another
type of mixed supersymmetry breaking transmission mechanism, gravity and gauge
mediation, with interesting phenomenological signatures.

%%%%%%%%%%%%%%%%%%%%%%%%%%%%%%%%%%%%%%%%%%%%%%%%%%%%%%
%%%%%%%%%%%%%%%%%%%%%%%%%%%%%%%%%%%%%%%%%%%%%%%%%%%%%%

\section{Gauge models of supersymmetry breaking and moduli stabilization}

Both main mechanisms of supersymmetry breaking, Fayet-Iliopoulos
(FI) \cite{fi} and O'Raifeartaigh (O'R) \cite{or} contain mass
scales, which fix the scale of supersymmetry breaking. If the
mediation of supersymmetry breaking is by Standard Model gauge
interactions, the corresponding mass scales are typically low. In
the case of the gravity mediation, they are typically at
intermediate energy scales. In both cases, a dynamical origin for
these values, small in Planck units, is needed. The traditional
viewpoint was to invoke some field theoretical nonperturbative
effects in a sector which dynamically breaks supersymmetry. With the
more recently studied stringy instantonic effects \cite{instantons},
there is the possibility of replacing field-theoretical
nonperturbative effects with stringy instanton effects which are
computable in string theory. These effects give mass scales of order
\begin{equation}
m_i \sim e ^{S_{\rm inst}} \quad , \quad {\rm where} \quad S_{inst}
= \sum_m c_m T_m \label{models01}
\end{equation}
is an instantonic action depending on moduli fields related to the
cycle(s) the instantonic brane is wrapping.  In all effective string
models we are aware, there are abelian gauge factors which are
gauged by some of the moduli fields, generating a Green-Schwarz
mechanism of anomaly cancelations. The crucial point we would like
to insist in what follows is that, while finding string effective
models reproducing at low-energy the basic features of the FI or the
O'R model in this framework is relatively easy, the dynamics of the
moduli fields $T_m$ set severe constraints on the transmission of
supersymmetry breaking. 

Let us start with a quick reminder of some basic results in ${\cal
N}=1$ supergravity with $U(1)$ gauged symmetries.
 As a starting point, let us recall that the scalar potential
in supergravity has the following
form:
\begin{equation}
\label{potential}
V = e^G (G^M ~G_M - 3) + {1 \over 2} \sum_A g_A^2 D_A^2 \ ,
\end{equation}
where
\begin{equation}
G \ = \ {\cal K} \ + \ \log |W|^2 \ ,
\end{equation}
and $ G_M = {\partial G \over \partial z^M}, $ where $z$ represents the
scalar part of a chiral superfield. The index $M$ runs over all the
chiral superfields present,  matter as well as hidden sector and/or
moduli fields. On the other hand, the auxiliary D-terms are given by
\begin{equation}
D_A = z^I T_{IJ}^A {\partial G \over \partial z^J} ~+~ \xi_A =
 {\bar z}^{\bar I} T_{IJ}^A {\partial G \over \partial
{\bar z}^{\bar J}}   ~+~ \xi_A \ , \label{s1}
\end{equation}
where $T^A$ represents the generators of the gauge group with index
$A$ and $\xi_A$ denotes the Fayet-Iliopoulos terms for the abelian
$U(1)$ factors. Note that the equality between the two last terms is
a straightforward consequence of the gauge invariance of the K\"ahler
potential. We take into account the fact that in string theory, the
Fayet-Iliopoulos terms are moduli-dependent.

In the presence of anomalous non-linearly realized abelian gauge symmetries
\begin{eqnarray}
&& \delta V_A \ = \ \Lambda_A +{\bar \Lambda}_A \quad , \quad
\delta z^i \ = \  2 X_i^a \ z^i \ \Lambda_A  \ , \nonumber \\
&& \delta T^{\alpha} \ = \ \delta_A^{\alpha} \ \Lambda_A \ , \label{models2}
\end{eqnarray}
the auxiliary D-terms are defined as
\begin{equation}
D_A \ = \ z^i \ X_i^A \ \partial_i G \ - \ {\delta_A^{\alpha} \over 2}
\ \partial_{\alpha} G \ . \label{DF1}
\end{equation}
The last term in (\ref{DF1}) is the moduli-dependent FI term
\begin{equation}
\xi_A \ = \  - \ {\delta_A^{\alpha} \over 2} \ \partial_{\alpha} G \
. \label{fi}
\end{equation}
In the following we will often use the standard definition of F-terms
\begin{equation}
F^i \ = \ e^{{\cal K} \over 2} \ {\cal K}^{i {\bar j}} \ \overline{D_j W}
\ ,
\end{equation}
where $D_j W = \partial_j W + W \partial_j K$,
such that (\ref{DF1}) becomes
\begin{equation}
m_{3/2} \ D_A \ = \ z^i \ X_i^A \ F_i \ - \ {\delta_A^{\alpha} \over
  2} \ F_{\alpha}  \ , \label{DF2}
\end{equation}
where as usual $m_{3/2} = \exp({\cal K}/2) W$. Eq. (\ref{DF2}) tells us
that the supersymmetric $D$ and the $F$ conditions are not
independent. Therefore in order to find supersymmetric solutions it
is enough to check the minimum independent number of SUSY
conditions.

Some comments are in order concerning moduli masses. In models with anomalous $U(1)$ and FI term
much larger than the gravitino mass, which is the case of interest in string theory and is the case considered
throughout this paper, there is always a charged field, of appropriate charge and large vev that compensates
to a good accuracy the FI term. This field is always heavier than the K\"ahler modulus transforming non-linearly
under $U(1)$, which has masses of the order (or slightly larger) than the gravitino mass. In the case of a twisted 
modulus, its mass is generically of the order of the string scale and therefore generically very heavy. Even in this case 
which suggests that the twisted modulus is irrelevant at low energy and therefore can integrated out, this has to be 
done with care. Indeed, as we will see in detail later on, supergravity approximation and gauge domination would 
require a very small vev for the twisted modulus. The explicit dynamics of the modulus, on the other hand, constrained 
by gauge invariance, often prefers large vev's.

%%%%%%%%%%%%%%%%%%%%%%%%%%%%%%%%%%%%%%%%%%%%%%%%%%%%%%%%%%

\subsection{Dominance of gauge mediation over gravity mediation}
In a generic model with (almost) Minkowski vacuum, and with
$F$-terms larger than the $D$-terms, the uplifting condition
requires a cancelation between the $F$-term contribution to the
vacuum energy and the SUGRA correction $-3m_{3/2}^2$. This 
implies that in a viable SUSY breaking minimum no $F$-term can
be much larger than $m_{3/2}$. On the other
hand, if SUSY is broken by some field S, coupled to the
messengers\footnote{We take $S$ to have standard K\"ahler
potential for simplicity.}, then dominance of gauge mediation over
gravity mediation requires
\begin{equation}
\frac{g^2}{16 \pi^2}
\frac{\langle F_S\rangle}{\langle S\rangle} \ \gg \ m_{3/2} \ .
\end{equation}
Thus, we have gauge mediation dominance only if  $\langle S\rangle \ll1$, i.e. if the messenger
masses $m\sim \langle S\rangle$ are very small in Planck units. Since
$F_S \sim \sqrt{3} m_{3/2} M_P$, we can define the ratio between
the gauge and gravity contributions to soft terms as
\begin{equation}
{M_{GM} \over m_{3/2} } \ 
\equiv \ g^2 \ \alpha \ \equiv {g^2 \over 8 \pi^2} \ \sqrt{3} \ N \ {M_P \over S} \ , \label{alpha} 
\end{equation}  
where $N$ is the number of messengers and $g$ a Standard Model gauge
coupling.  
In a (minimal) model with only one field $S$ charged under some anomalous $U(1)_X$,
the value of $S$ is roughly set by the value of the FI terms
$\xi$, so that we can have a small $\langle S\rangle$ only in
the presence of a small $\xi$. By fixing $\xi$ to be very small,
we are {\it always} enhancing the contribution of gauge mediation
over the gravity mediation.

On the other hand, as explained above in (\ref{fi}), $\xi$ is
moduli-dependent and its value is fixed by the dynamics. We cannot assume that $\xi$ is fixed 
by some independent dynamics ``at high scale'',
since such a dynamics involves moduli stabilization, that affects the form of
the $D$-term potential. The problems stated here
could perhaps be avoided but only  in a complicated generalization of the minimal constructions, including more than one modulus field.

In what follows we show in detail how the minimal setups fail in achieving gauge mediation
dominance, due to the large vev's of $S \gsim 10^{-3}$ 
and show a model where only a marginal dominance $\alpha \lsim 100$ is
allowed for a large number of messengers. Then,  as discussed later, experimental
constraints on superpartner  masses imply that there is lower limit on the gravitino mass
$m_{3/2} \gsim 50-100$ GeV. 
The resulting, {\it hybrid} phenomenology, will be the main subject
of section 5.

%%%%%%%%%%%%%%%%%%%%%%%%%%%%%%%%%%%%%%%%%%%%%%%%%%%%%%%%%%

\subsection{Gauged Polony model and SUSY vacua}

The simplest model falling in the description given above contains
one single field $S$ with a linear superpotential term $W\supset
\lambda S $ breaking SUSY. In order to have a viable model,
$\lambda$ should be small, thus sourced by some instantonic effects
involving, in a minimal string setup, one modulus field $T$:
$\lambda \sim e^{-\alpha T}$. Assuming $S$ to be charged under a
$U(1)_X$, as it is always the case in a string model, gauge invariance forces $T$
to have an anomalous variation. Thus, under the gauge transformation
$V\rightarrow V+\Lambda+\bar \Lambda$, we have $S\rightarrow S e^{-2
q_S \Lambda}$ and $T\rightarrow T+\delta_{GS} \Lambda$, $q_S$ being
the $S$ $U(1)_X$ charge. We use the notation $\delta_{GS}$ since we
expect the anomalous $T$ variation\footnote{If we force $T=T_0$ this is the model recently
studied in \cite{hv}}  to be linked with an anomaly
cancellation in a string models, via the Green-Schwarz mechanism,
i.e. we expect $T$ to be part of some gauge kinetic function. Taking
$q_S=-1$ we have
\begin{eqnarray}
&& W \ = \ W_0 \ + \ a e ^{- {2 T \over \delta_{GS}} } \ S \ , \ \nonumber \\
&& {\cal K} \ = \ S^{\dagger} e ^{-2 V} S \ + \  K (T +
{\bar T} - \delta_{GS} V) \ ,  \nonumber \\
&& D \ = \ - |S|^2 \ - \ {\delta_{GS} \over 2} \ \partial_T K \ , 
\label{models4}
\end{eqnarray}
where the last term in the $D$-term is the field dependent FI term
due to the anomalous $T$ variation. In this model, the relation
(\ref{DF2}) between $F$-terms and $D$-terms can be rewritten as
\begin{eqnarray}
S \ F_S  \ + \ \frac{\delta_{GS}}{2} \ F_T  \ = \ - \ D \ m_{3/2} 
\end{eqnarray}
and a SUSY minimum is ensured if
\begin{equation}
D = - |S|^2 - \frac{\delta_{GS}}{2} \partial_T K = 0, \quad
e^{-{\cal K}/2}F_S=a e^{-\frac{2 T}{\delta_{GS}}} +\bar S W = 0 \ .
\end{equation}
These two equations generically have solution, but the specific
details depend on the form of $K (T +{\bar T} - \delta_{GS} V)$.
In what follows we consider some cases of particular relevance in a string theory setup:
first we study the case in which $T$ is a standard K\"ahler modulus, with logarithmic potential,
then we consider the case of a twisted modulus, with polynomial potential. Finally we consider
the generalization to the case in which the sector of moduli contains more than one modulus
field.

Our aim is to show that under very general assumptions, it is difficult to have dominance
of gauge mediation over gravity mediation, since no viable minimum is present with $\langle S\rangle \ll 1$.

\paragraph{T as a standard K\"ahler modulus}\mbox{}\\
In this case
\begin{equation}
K = - 3 \, \log (T +{\bar T} - \delta_{GS} V) \ ,
\end{equation}
thus the SUSY conditions can be rewritten, neglecting the imaginary parts of $S$ and $T$,  as
\begin{eqnarray}
D=-|S|^2+\frac{3\delta_{GS}}{2(T+\bar T)}=0,\quad e^{-{\cal K}/2}F_S=a e^{-\frac{3}{2 |S|^2}}
(1+|S|^2) +\bar S W_0 = 0 \ .
\end{eqnarray}
The first equation fixes $T$ as a function of $S$, then $T$ was
substituted in the condition $F_S=0$. The second equation has a
solution for $|S|^2\rightarrow 0$, corresponding to the runaway
$T\rightarrow \infty$. In the case $W_0 \ll a$  there is another SUSY
solution for $\langle S\rangle \ll 1$, and $T \gg 1$.
Since typically $a\sim {\cal O}(1)$, we have that in such a model,
for essentially all the reasonable values of $W_0$, there is a SUSY
minimum in a viable region for $\langle S\rangle$ and $\langle T\rangle$ and therefore this model
does not break SUSY.

Moreover, even in case a metastable SUSY breaking vacuum is found,
we can infer, on a very general basis, that it is impossible to obtain a pure gauge mediation scenario.
After coupling $S$ to some messenger fields, we immediately find that the condition $\langle S \rangle\ll 1$
is incompatible with the original supergravity dynamics. In fact, from $D$-flatness one finds that
the instantonic effect is of order
\begin{equation}
e^{- {2 T \over \delta_{GS}}} \ = \  e^{- {3 \over 2 S^2}} \ .
\label{models7}
\end{equation}
In other words,  the requirement of a very small FI term kills the SUSY breaking effects.

\paragraph{T as a twisted K\"ahler modulus}\mbox{}\\
In this case we expect $T\ll1$, thus, the non-perturbative effects are suppressed only
assuming the gauge kinetic function to depend on some extra modulus $M$, having
no transformation under the anomalous $U(1)_X$.
Since $M$ is invariant under a $U(1)_X$ transformation, it does not enter in the $D$-term,
its dynamics is not constrained by gauge invariance, and it is reasonable to assume
some hidden dynamics to stabilize it at high scale and large vev.
In this way, the net effect of $M$ is a small prefactor $a$ in front of the instantonic
superpotential term.
Given the K\"ahler potential
\begin{equation}
K=\frac{1}{2}(T +{\bar T} - \delta_{GS} V)^2 \ + \ S^{\dagger} e ^{-2 V} S \ ,
\end{equation}
the SUSY conditions can be rewritten as
\begin{eqnarray}
D=-|S|^2-\frac{\delta_{GS}}{2}(T+\bar T)=0,\quad
e^{-{\cal K}/2}F_S=a e^{\frac{2 |S|^2}{\delta_{GS}^2}} (1+|S|^2) +\bar  S W_0=0,
\end{eqnarray}
where we assume $\delta_{GS}<0$, and we neglected the imaginary parts of  $T$ and $S$.
Thus, under the assumption $S\ll1$ we find viable SUSY solutions only for $a\ll W_0$: $S \sim -a/W_0$.
In other words, in case  $a\ll W_0$  a SUSY solution is present at $S\sim -a/W_0$. Moreover, in the limit 
$a\ll W_0$ it's easy to check that any non-SUSY minimum eventually present,
with $S\ll 1$,  cannot be uplifted, since $F_S,\,F_T\ll m_{3/2}$.
Thus, the only viable regime of the parameters is $a\gg W_0$,
but in such a case the requirement $S\ll 1$ forces $F_S\gg m_{3/2}$. This implies that no cancelation between
$F_S$ and $W$ is possible, so that the cosmological constant is set by the $F_S$ value to be larger than
the gravitino mass.
Of course, one could in principle hope a viable SUSY breaking minimum to be present in case
$a\sim W_0$, where the problems raised above could be absent. On the other hand, it is unreasonable to
expect such a minimum for small values of $\langle S\rangle$.
Indeed, assuming $a\ll1$ we see that the $D$-term contribution will fix $S$ close
to the modulus vev (or vice-versa), plus $O(a)$ corrections. Replacing this in the other equation we essentially get
just an algebraic function with order one parameters, multiplied by an overall energy scale $a^2$.
Thus, no extrema exist at small $S$ (as it can also be checked by direct inspection).

\paragraph{Extensions to a generic number of moduli fields}\mbox{}\\
A natural extension is to consider more than one modulus to be charged under the
anomalous $U(1)_X$.
Let us consider $n$ moduli $T_i$, $i=1,\dots n$ with K\"ahler potential\footnote{
In the case in which the moduli are all ``twisted'' we get results similar to the
single twisted modulus case.}
\begin{equation}
K \ = \ - \ \sum_i p_i \, \log (T_i+\bar T_i-\delta_i V) \ ,
\end{equation}
the superpotential
\begin{equation}
W \ = \ W _0 \ + \ a e^{-\sum_{i} b_i T_i} S
\end{equation}
with $\sum b_i \delta_i=2$ and the D-term
\begin{equation}
D = - |S|^2+\sum_i\frac{p_i\delta_i}{2}\frac{1}{T_i+\bar T_i} \ ,
\end{equation}
while the $F$-terms are
\begin{eqnarray}
&&e^{-{\cal K}/2}F_S = \bar S W+a e^{-\sum_i b_i T_i} \ , \nonumber \\
&&e^{-{\cal K}/2}F_{T_i} = - \frac{p_i}{T_i+\bar T_i} W \ - \ a b_i \ e^{-\sum_i b_i T_i}S \ .
\end{eqnarray}
A SUSY solution is present if we have all F-terms equal to zero.
Thus, we can replace the condition $F_S=0$ in the other conditions and find
\begin{eqnarray}
\frac{p_i}{T_i+\bar T_i} \ = \ b_i \ |S|^2 \ .
\end{eqnarray}
These conditions, that also imply $D=0$, fix the real parts of $T_i$ as
functions of $S$. Thus, neglecting the $T_i$ phases, we can replace them
in the eq. $F_S=0$ to find
\begin{equation}
\label{constr}
\bar S \ W \ + \ a e^{-\frac{\sum_i p_i}{2 |S|^2}} = 0 \ ,
\end{equation}
that is, de facto, {\it the same} condition found in the single (untwisted) modulus case,
having solution in all the reasonable regimes for $W$, and with the same (negative)
implications for what concerns the issue of gauge mediation dominance.

A natural ``complication'' arising in the presence of more than one modulus is
the presence of extra instantonic effects, so that
\begin{equation}
W\supset \sum_j e^{-\sum_i \alpha^j_i T_i},
\end{equation}
with $\sum\alpha^j_i\delta_i=0$ for each $j$, due to gauge invariance.
Such new terms can have an important r\^ole in the stabilization of the moduli
fields, but it is clear that the F-term equations will be generically solvable,
and we expect a generic model to have a SUSY minimum in some viable parameters
region. Of course, in very specific models the minima could be located at unacceptable
values of the moduli fields, and extra SUSY breaking minima could be present.
On the other hand, in such a case it would be precisely the moduli sector the main actor
and the most interesting part of the whole SUSY breaking mechanism.
In such a generalization one could hope to avoid the constraint
of Eq.~(\ref{constr}), and fix the moduli so to have a small induced FI term
without destroying the instantonic effect breaking SUSY. Such a model would
be a good candidate for a model realizing gauge mediation in a realistic
string model.

%%%%%%%%%%%%%%%%%%%%%%%%%%%%%%%%%%%%%%%%%%%%%%%%%%%%%%%%%%

\subsection{Fayet-O'Raifeartaigh models}

The simplest model in this class is again based on a $U(1)_X$ gauge
symmetry with a O'R like superpotential
\begin{eqnarray}
&& W \ = \ W_0 \ + \phi_{--} (\lambda_1 \phi_{+}^2 - m^2 e^{-{4 T
\over \delta_{GS}}})
\ + \ \lambda_2 \ \chi_{--} \phi_{+}^2 \ , \ \nonumber \\
&& {\cal K} \ = \ \phi_{+}^{\dagger} e ^{2 V} \phi_{+} +
\phi_{--}^{\dagger} e ^{-4 V} \phi_{--} + \chi_{--}^{\dagger} e ^{-4
  V} \chi_{--} +  K (T + {\bar T} - \delta_{GS} V) \ ,  \nonumber \\
&& D \ = \ -2 \ |\phi_{--}|^2 \ - \ 2 \ |\chi_{--}|^2 + |\phi_{+}|^2 -
{\delta_{GS} \over 2} \ \partial_T K  \ . \label{models9}
\end{eqnarray}
It is easy to check that in the supersymmetry breaking vacuum, we have
\begin{eqnarray}
&& {\bar F}_{\phi_{+}} = 0 \qquad \qquad {\bar F}_{\phi_{--}} =  
- \frac{\lambda_2}{\lambda_1}{\bar F}_{\chi_{--}}= 
- \frac{\lambda_2^2 m_0^2}{\lambda_1^2 + \lambda_2^2}\qquad \qquad  D=0  \\
&& |\phi_{+}|^2 = \left| \frac{\lambda_1 m_0^2}{\lambda_1^2 + \lambda_2^2}\right| \qquad
\left| \phi_{--} \right|^2 = \frac{\lambda_2^2}{\lambda_1^2}\left|\chi_{--} \right|^2 = 
\frac{1}{2} \left( \xi_{FI}^2 + \left| \frac{\lambda_1 m_0^2}{\lambda_1^2 + \lambda_2^2}\right|\right) \frac{\lambda_2}{\lambda_1^2 + \lambda_2^2}
\nonumber 
\end{eqnarray}
where we have defined $\xi_{FI}^2 = -{\delta_{GS} \over 2} \ \partial_T K  $ and
$m_0^2 = \exp(K/2) m^2 \exp (-4 T / \delta_{GS})$. 
About the chance of having gauge mediation dominance, we observe that
\begin{equation}
\left|\frac{F_{\phi_{--}}}{\phi_{--}}\right|^2 =\left|\frac{F_{\chi_{--}}}{\chi_{--}}\right|^2 =
 \frac{2 \lambda_2^2 m_0^4}{\left( \lambda_1^2 + \lambda_2^2 \right)} \frac{1}{\xi_{FI}^2 +
 \left|\frac{\lambda_1 m^2}{(\lambda_1^2 + \lambda_2^2)}\right|} < \frac{6 m_{3/2}^2}{\xi_{FI}^2} \ .
\end{equation}
Once again, the only possibility to have pure gauge mediation requires $\xi_{FI}^2 \ll 1$, but this is incompatible with
moduli stabilization, at least in case we have a single modulus with logarithmic K\"ahler potential.

\subsection{Fayet-Iliopoulos and Fayet-Polony models}
The simplest generalization of the basic Fayet-Iliopoulos model is
described by an $U(1)_X$ gauge symmetry and two charged fields
$\phi_{\pm}$, plus a mass term:
\begin{eqnarray}
&& W \ = \ W_0 \ + \ e ^{- M } \phi_{+} \phi_{-} \ , \ \nonumber \\
&& {\cal K} \ = \ \phi_{+}^{\dagger} e ^{2 V} \phi_{+} +
\phi_{-}^{\dagger} e ^{-2 V} \phi_{-} + K_0 (M +{\bar M}) + K (T +
{\bar T} - \delta_{GS} V) \ ,  \nonumber \\
&& D \ = \ |\phi_{+}|^2- |\phi_{-}|^2 - {\delta_{GS} \over 2}
\partial_T K \ , \label{models1}
\end{eqnarray}
where the last term is the field-dependent FI parameter.
Under gauge
transformations, the various fields transform as
\begin{eqnarray}
&& V \rightarrow V + \Lambda +{\bar \Lambda} \quad , \quad
\phi_{\pm} \rightarrow e^{ \mp 2 \Lambda} \phi_{\pm} \ , \nonumber \\
&& T \rightarrow T +  \delta_{GS} \ \Lambda \ . \label{models02}
\end{eqnarray}
Since $T$ is (nonlinearly) charged under $U(1)_X$ whereas the
operator $\phi_{+} \phi_{-}$ is neutral, the instantonic action
generating the mass term $m \sim e^{- M}$ cannot depend on $T$, but
it should depend on another, $U(1)_X$ neutral modulus (or linear
combination of moduli) $M$. The constant $W_0$ in (\ref{models1})
was added for two reasons. First, it will help to adjust the
cosmological constant to zero in a supergravity framework. Secondly,
as well-known, it also plays an instrumental role in moduli
stabilization. This model has the feature that, {\it if} the moduli
fields are assumed to be stabilized, in the rigid (global) limit
it reduces to the FI model of supersymmetry breaking. From the point
of view of moduli stabilization, however, the model (\ref{models1})
is not very satisfactory, since there are two moduli fields, $M$ and
$T$ to stabilize. Whereas $M$ could be stabilized by string theory
fluxes, along the lines of \cite{Kachru:2003aw} and simultaneously
generating $W_0$, $T$ cannot be stabilized without further dynamics. \\
  A more viable model is what we could call a Fayet-Polony model, with the same K\"ahler potential
and D-term as in (\ref{models1}), whereas the superpotential is
\begin{equation}
W \ = \ W_0 \ + \ e ^{- M } \phi_{+} \phi_{-}
\ + \ e ^{- 2 qT / \delta_{GS}} \phi_{-}^{q} \ . \label{models3}
\end{equation}
The powers $q$ are selected such that $2 q T / \delta_{GS}$
represents an integer of the instanton action. By assuming as above
that $M$ is stabilized by fluxes (or that the mass $m= \exp(-M)$ is generated by
field-theory nonperturbative dynamics), this model was analyzed in
detail in \cite{dmpr} for the case of the volume-type $T$
modulus
\begin{equation}
K \ = \ - 3 \ \log (T+{\bar T} - \delta_{GS} V) \ . \label{models03}
\end{equation}
Strictly speaking, only $q=1$ corresponds to a linear, Polony-like
term, but since $\phi_-$ tends to compensate the FI term and to get
a large vev, the dynamics of the model is not qualitatively
different for $q=1$ and $q \not=1$.
It is conceptually transparent that in a model like  (\ref{models3})
moduli stabilization cannot be ignored. First of all, the modulus
mass $T$ has a mass that, for large FI term is parametrically
smaller than the mass of $\phi_{-}$ so cannot be consistently
integrated out.
Secondly, even if this could be possible, freezing out $T$ in
(\ref{models3}) gives a gauge non-invariant lagrangian. The
computation of any physical quantity done in this way differs
considerably from the computation done by keeping the modulus
dynamics. In the
following sections of the present paper we show that
only a moderate dominance is possible
in very specific regions of the parameter space.

%%%%%%%%%%%%%%%%%%%%%%%%%%%%%%%%%%%%%%%%%%%%%%%%%%%%%%%%%%%

\subsubsection{SUSY minima in the Fayet-Polony model}
We think the modulus $M$ stabilized at high energy \footnote{Note that this is possible only because the modulus $M$
can decouple from the dynamics of the $U(1)_X$ sector.} and replace it with a constant $m=e^{-M}$ in the model (\ref{models3}).
A complete analysis of the vacuum structure of this model has already been done in \cite{dmpr} and extended in the next sections.
Here we explicitly show the role of the parameter $m$ in determining the presence or the absence of the supersymmetric solution.
For example, the relation between F-terms and D-term in the case $q=1$, reads
\begin{equation}
\phi_{+} F_{+} - \phi_{-} F_{-} - \frac{\delta_{GS}}{2} F_T \ = \ D~m_{3/2} \ .
\end{equation}
It is then enough to search for solutions of the following equations
\begin{eqnarray}
e^{-{\cal K}/2}F_{+} &=& m \phi_{-} + {\bar \phi}_{+} W = 0 \ , \nonumber \\
e^{-{\cal K}/2}F_{-} &=& m \phi_{+} + a e^{-2T/\delta_{GS}} + {\bar \phi}_{-} W = 0 \ , \nonumber \\
D &=& |\phi_{+}|^2 - |\phi_{-}|^2 -\frac{\delta_{GS}}{2} \partial_T K = 0 \ .
\end{eqnarray}
It is straightforward to see that the model has a possible SUSY vacuum for
\begin{eqnarray}
\phi_{+} &=& - \frac{{\bar m}}{{\bar W}} {\bar \phi}_{-} \ , \nonumber \\
\phi_{-} &=& \frac{a e^{-2{\bar T}/\delta_{GS}} ~W}{|m|^2 - |W|^2} \ , \nonumber \\
\partial_{T} K &=& \frac{2}{\delta_{GS}} \frac{|a e^{-2{\bar T}/\delta_{GS}}|^2}{\left(|m|^2 - |W|^2\right)}
\ .
\end{eqnarray}
However, the last equation shows that, depending on the sign of $\partial_{T} K$, a solution is allowed in different regions
of the parameters space. In particular, focusing on the case of a standard K\"ahler modulus $T$ (\ref{models03}), the equation
\begin{equation}
- \frac{3}{T + {\bar T}} = \frac{2}{\delta_{GS}} \frac{|a e^{-2{\bar T}/\delta_{GS}}|^2}{\left(|m|^2 - |W|^2\right)}
\end{equation}
admits sensible physical solutions only for $m \ll W$. This agrees with the limit $m \to 0$, recovering the gauged Polony
model discussed above. However, choosing $m \gg W_{0}$ and requiring $\phi_{+}, \phi_{-} \ll 1$, is enough to assure
that no SUSY solution exists: this is actually the case studied 
in the rest of the paper. Note that the condition $m \gg W_{0}$ is also the same assuring the uplifting of the SUSY breaking vacuum.\\
The crucial point is that now, since the dynamics of the modulus $T$ is really decoupled from the $U(1)_X$ sector, with
the new parameter $m$ in the model we can create a hierarchy in the mass scales and SUSY can be spontaneously broken.

\subsubsection{Gauge dominance in the Fayet-Polony model}
As we will show later in Sect. 4, in the metastable SUSY breaking minimum
\begin{eqnarray}
F_{\phi_+} \sim m_{3/2}\ ,\quad F_{\phi_-} \sim m_{3/2}\,\, \xi_{FI}\\
\phi_-\sim \xi_{FI}\ , \quad \phi_+ \sim \phi_-^2 \sim \xi_{FI}^2,
\end{eqnarray}
with 
\begin{equation}
\xi_{FI}^2 \equiv -\frac{\delta_{GS}}{2}\partial_T K 
\end{equation}
so that
\begin{equation}
\frac{F_-}{\phi_-}\sim m_{3/2}\ ,\quad \frac{F_+}{\phi_+}\sim \frac{m_{3/2}}{\xi_{FI}^2}.
\end{equation}
Thus, the SUSY breaking field is $\phi_+$, the field $S$ using the notation introduced before.
What is specific of our model is that $S$ is not set to $\xi_{FI}$, but rather to $\xi_{FI}^2$, 
and even in presence of a ``not-so-small'' FI term we have a reasonable enhancement
of gauge mediation. In our case the minimization fixes $\xi_{FI} \lesssim 10^{-1}$, and 
some dominance of gauge mediation is possible, as we show later in some more detail.

From this we see that the last model is qualitatively better than the others discussed
in the rest of the section: it is the simplest model with one modulus
where it is possible to have a (meta)stable SUSY breaking
vacuum, with the gravitino mass much larger than the cosmological constant, tunable to small
values, for vev's of the fields falling in the region where the SUGRA
regime can be trusted. This is partially since, after modulus
stabilization, it seems hard to get a dynamically small FI term.  Our
results suggest that strong dominance of gauge mediation in the
present setup necessarily request a multi-moduli setup with a
non-trivial dynamics, which is beyond the goals of the present work.     
In what follows, after a brief summary of general facts about gauge mediation of SUSY breaking in the
presence of non-zero D-terms (such as in our model), we turn to a detailed study of the microscopic
and phenomenological properties of this model.

%%%%%%%%%%%%%%%%%%%%%%%%%%%%%%%%%%%%%%%%%%%%%%%%%%%%%%%%%%%
%%%%%%%%%%%%%%%%%%%%%%%%%%%%%%%%%%%%%%%%%%%%%%%%%%%%%%%%%%%

\section{Gauge mediation: standard and non-standard contributions }\label{motivations}

Gauge mediation can be defined perturbatively as containing some messenger fields $M_i, {
\tilde M}_i$, vector-like with respect to the SM gauge group, coupling to some chiral SUSY breaking fields $S_a$,
and eventually charged under the possible $U(1)_X$ factors present in the SUSY breaking sector. The messengers
are defined to couple directly only to $S_a$ via couplings as
\begin{equation}
W_m \ = \ M_i \ (\lambda_{ija} S_a + \mu_{ij}) \ {\tilde M}_j \ , \label{g1}
\end{equation}
where $\lambda_{ija}$ and  $\mu_{ij}$ are fixed by hidden sector $U(1)_X$ gauge invariance constraints. It is convenient
in what follows to consider matrices in the messenger space ${\hat \lambda}_a$ and $\mu$.
Non-renormalizable couplings to $S_a$ are also possible, but they do not change the conclusions of the present
discussion. Let us start the discussion for simplicity with one field
$S$, and $\mu=0$.
The contributions to the soft terms from the
messengers are encoded in their mass matrix and
in particular in the value of its eigenvalues and supertrace. The
scalar mass matrix for a couple of messengers
generically coupled to a superfield $S$ with a coupling $\lambda$, and
with charges $q, {\tilde q}$ under
$U(1)_X$ is

\begin{equation} \label{matrixmess}
{\cal M}_{0}^2=
\left(
\begin{array}{cc}
\lambda^2 \langle S \rangle^2 + q g_X^2 D & \lambda F_S \\
\lambda F_S & \lambda^2 \langle S \rangle^2 + {\tilde q} g_X^2 D
\end{array} \right)
\end{equation}

\noindent
where $D$ is the D-term of $U(1)_X$, $g_X^2$ its coupling constant and $F_S$ the auxiliary field of $S$.

It is well-known (but sometimes overlooked) that in the presence of $D$-term contributions in the
hidden sector, that are generic in string theory constructions, there are some constraints and ingredients
to take into account:

\begin{itemize}

\item The absence of the one-loop induced Fayet-Iliopoulos term $\xi_Y$ for the hypercharge imposes the condition \cite{dnns}
\begin{equation}
\xi_Y \sim Tr \biggl( Y \int {d^4 k \over (2 \pi)^4} {1 \over k^2 +
  {\cal M}_0^2} \biggr) = 0 \ , \label{g2}
\end{equation}
where ${\cal M}_0^2$ is the scalar messenger mass matrix.
Notice that the condition (\ref{g2}) is
stronger than the absence of the logarithmically divergent
piece\footnote{The quadratic divergence cancels since $Tr Y=0$.}
\begin{equation}
(Tr \ Y {\cal M}_0^2)_{\rm mess} \ \sim \ (Tr \ Y X) \ \langle D \rangle
\ = \ 0 \ ,
\end{equation}
where $X$ is the  generator of the hidden sector $U(1)_X$, that is equivalent to the absence of mixing between hypercharge and
$U(1)_X$. A sufficient (but not necessary) condition which satisfies (\ref{g2}) that we will use in the
rest of the paper, is to consider vector-like (with respect to the SM
gauge group) messenger fields with equal $U(1)_X$ charges $q = {\tilde q}$
\begin{eqnarray}
&& \qquad \qquad U(1)_Y \qquad \qquad U(1)_X \qquad  \nonumber \\
&& M \qquad \quad \ \ \ y \qquad \qquad \qquad q \qquad \qquad  \nonumber \\
&& {\tilde M} \qquad \quad -y \qquad \qquad \qquad q \qquad \qquad \label{messcharges}
\end{eqnarray}
where we displayed only the abelian charges of the messenger fields
$M,{\tilde M}$.  Notice that having vector-like messenger fields with
respect to all gauge groups, i.e. charges $(y,q)$ for $M$ and
$(-y,-q)$ for ${\tilde M}$ {\it does} generate a FI term for the
hypercharge which phenomenologically, if non-zero, has to be very
small. Some recently proposed string models with supersymmetry
breaking and gauge mediation \cite{aks} fall into this
category and could therefore be phenomenologically problematic.

\item If $(Str \ {\cal M}^2)_{\rm mess} \not= 0$, there are new contributions to the MSSM scalar masses
(but not to the gaugino masses) \cite{Poppitz:1996xw,dmpr}. They can generate
phenomenological problems \cite{Poppitz:1996xw}
or, on the contrary, in
a well-defined theory containing gravity, can generate an original compressed low-energy spectrum, with squarks
lighter than sleptons at high energy \cite{dmpr}. These new terms are proportional to\footnote{
Notice that $\sum_i Str \ {\cal M}_{mess,\,i}^2\equiv Str \ {\cal M}_{mess}^2=0$ is not enough to guarantee
the absence of non-standard contribution.}
\begin{equation}
\sum_i (Str {\cal M}^2)_{{\rm mess},i} \ \log {\Lambda^2 \over
  m_{f,i}^2} 
\sim \sum_i \ (Tr \ X_i) \ \log {\Lambda^2 \over
  m_{f,i}^2} \ g_X^2 \langle D \rangle \  \ , \label{g3}
\end{equation}
where $i$ labels vector-like messengers and $m_{f,i}$ are the
messenger fermionic masses.    
 More details about this formulae can be found in the appendix A2.
\end{itemize}

For one messenger pair of charge $q = {\tilde q}$, the two eigenvalues of (\ref{matrixmess}) are
given by:
\begin{equation}
m_-^2= \left[(\lambda \langle S \rangle)^2 + q g_X^2 D\right] - \lambda
F_S \ , ~~~~~~~
m_+^2= \left[(\lambda \langle S \rangle)^2 + q g_X^2 D\right] + \lambda
F_S \ , \label{messmasses}
\end{equation}

\noindent whereas the fermion mass is given by:

\begin{equation}
m_f \ = \ \lambda \langle S \rangle \ .
\end{equation}

The supertrace is then
\begin{equation}
(Str M^2)_{\rm mess.} \ = \ 2 q \ g_X^2 \ D \ \not = \ 0 \ . \label{mess1}
\end{equation}

By standard gauge-mediation type diagrams, gaugino masses are induced
at one-loop, whereas scalar masses are induced at two-loops. However, as explained above, in the presence of a non-vanishing supertrace for the messengers,
the computation of the scalar masses is slightly different compared to
the standard gauge-mediation models \cite{Poppitz:1996xw}. In particular the result is not anymore UV
finite, there is a logarithmically divergent term.

Whenever we are interested in a predominantly standard gauge mediation spectrum, in addition to the well-known condition
\begin{equation}
M_{GM} 
\gg m_{3/2} \label{g4}
\end{equation}
where $M_{GM} $ is the typical scale of the soft terms in standard gauge mediation, the vanishing of the two additional contributions (\ref{g2}),(\ref{g3}) has to be imposed.
On the more quantitative level, standard gauge mediation contributions dominate over non-standard ones (\ref{g3})
for
\begin{equation}
M_{GM}^2 \ \gg \  (Tr \ X) \ \langle D \rangle \ .  \label{g5}
\end{equation}

Whereas a small value of the induced FI term for the hypercharge (\ref{g2}) and some of  non-standard contributions (\ref{g3}) can be
allowed, their complete absence entails the following simple constraints
\begin{equation}
\sum_i \left(Tr \ X_i\right) \ \log \frac{\Lambda^2}{m_{f,i}^2} \ = \ 0 \quad , \quad    Tr \biggl( Y \int {d^4 k \over (2 \pi)^4} {1 \over k^2 +
  {\cal M}_0^2} \biggr) = 0 \ . \label{g6}
\end{equation}
The first equation reduces to $Tr X =0$ in the limit of equal messenger masses.

%%%%%%%%%%%%%%%%%%%%%%%%%%%%%%%%%%%%%%%%%%%%%
%%%%%%%%%%%%%%%%%%%%%%%%%%%%%%%%%%%%%%%%%%%%%
\section{A class of  hybrid models}\label{micro}

We have constructed in Section 2 a model
breaking supersymmetry in the ground state, with hierarchically
small SUSY breaking scale and uplift of vacuum energy to zero. Since
we are mainly interested in the gauge versus gravity mediation
transmission, we add now the most general messenger sector. We
exclude $U(1)_X$ charges for MSSM fields, since otherwise large soft masses are generated through
the D-term potential\footnote{In principle one possibility is to
give a charge to the first two generations only, in order to keep
some light superpartners and to minimize the electroweak fine-tuning
\cite{ckn}. However, the large hierarchy between these two
generations of squarks and the third one can generate other problems
like for example a tachyonic direction for the third generation
after the RGE flow towards low-energy \cite{ahm}.}. 
In  \cite{dmpr}, we considered only a minimal 
messenger sector, non-chiral with respect to the MSSM gauge group, and
having positive $U(1)_X$ charge. In that case, the phenomenology of
the model was characterized by a strong competition between
modulus/gravity mediation and non-standard GMSB. In the present generalization we enlarge the standard
GMSB contribution, and make (marginal) contact with models of pure
GMSB.

The fields that are relevant in our construction are $T$, $\Phi_-$
and $\Phi_+$, as introduced above, and a set of $N_*$ messengers
($\widetilde M_{(*)i}$) $M_{(*)i}$ in the (anti)fundamental
representation of $SU(5)$, with $*=$ positive ($n_i/2$), negative
($-m_i/2$), and zero $U(1)_X$ charge. We consider $T$ to have
  K\"ahler potential $-3 \log(T+\overline T-\delta_{GS} V)$, and all the other fields to have canonical kinetic terms.
Anomaly cancelation sets a mild constraint on the $U(1)_X$ charges. The exact constraint depend on the details of the
MSSM gauge kinetic functions $f_a = {\rm const} + c_a \ T$. In what follows we take $c_a=c$ in order to obtain unification,
and so the sign of the difference
between the number of the positively and negatively charged
messengers is fixed by 
\begin{equation}
\label{cdef}
\Sigma_{i=1}^{N_+} n_i - \Sigma_{j=1}^{N_-} m_j \ \sim \ c \ .
\end{equation}

Working in Planck units, we introduce the superpotential
 \begin{eqnarray} \label{superpotential}
&& W = W_0 + a e^{-b T} \Phi_{-}^q \left( \Phi_{+} \Phi_{-} \right)^p + m \Phi_+ \Phi_- \nonumber \\
&& \qquad + \Sigma_{i=1}^{N_+}  \lambda_{(+)i} \Phi_{-}^{n_i} M_{(+)i} \widetilde{M}_{(+)i} \nonumber \\
&& \qquad+  \Sigma_{j=1}^{N_-}  \lambda_{(-)j} \Phi_{+}^{m_j} M_{(-)j} \widetilde{M}_{(-)j} \nonumber \\
&& \qquad +  \Sigma_{r=1}^{N_0}  \left[ \mu_r + \lambda_{0r} \left( \Phi_{+} \Phi_{-} \right)^{l_r}\right]  M_{(0)i}
\widetilde{M}_{(0)i}
\end{eqnarray}
In what follows we briefly motivate the superpotential terms
introduced above, and explain why extra terms, allowed by the
symmetries, can be neglected. In detail, $W_0$ is what remains after
integrating out the moduli stabilized at high energy (e.g. the
complex structure moduli in a type IIB flux compactification), and
can be (unnaturally) small with respect to the Planck scale; the
term $e^{-b T} \Phi_{-}^q \left( \Phi_{+} \Phi_{-} \right)^p $ is
due to a condensating sector coupled to the fields $\Phi_-$ and
$\Phi_+$, after hidden sector meson fields have been integrated out.
In principle we should include also terms like $ \eta_n \left(
\Phi_{+} \Phi_{-} \right)^{n}$ with $n>1$.  However, we assume them
to be irrelevant, as it is the case if their origin is allowed only
at nonperturbative level for $n < n_0$ due to some discrete
symmetry. Indeed the small
value of the parameter $m$, needed for the stabilization and
uplifting procedure, is naturally explained in this framework.
Nonetheless, if $n_0$ is big enough \footnote{Roughly speaking, this
value will be greater than $log(m) / log \left( \langle\Phi_+
\rangle~\langle \Phi_-\rangle\right)$. } the contributions from this
kind of term for $n \ge n_0$ will be suppressed due to the smallness
of $\langle \Phi_+ \Phi_- \rangle^n$. Finally, we avoid also gauge
invariant terms coupling the messengers with negative charge to the
modulus, as for example $e^{-b T} M_{-}^{\frac{q}{2}}
\tilde{M}_{-}^{\frac{q}{2}}$, since they do not change neither the
minimizations
procedure nor the phenomenological results.\\
Mixing terms between messengers with opposite charges, like for example
\begin{equation}
\left[ \lambda \left( \Phi_+ \Phi_- \right)^{k} -\mu \right] \left( M_{-} \tilde{M}_{+} + M_{+} \tilde{M}_{-} \right),
\end{equation}
could be more problematic and we discuss it in the section (\ref{messengers}).\\
As we will see in the rest of the section, the presence of the new terms in the superpotential with respect to the
case \cite{dmpr} implies different problems and possibilities, concerning the stability of the phenomenologically
interesting vacuum and the mass spectra.

%%%%%%%%%%%%%%%%%%%%%%%%%%%%%%%%%%%%%%%

\subsection{Vacuum structure of the model}

The potential of the model is computed given: (i) the D-term potential
\begin{eqnarray}
\nonumber
V_D=\frac{4\pi}{T+\overline{T}}
&&\hspace{-20pt}
\left[|\Phi_+|^2-|\Phi_-|^2 + \frac{1}{2} \sum_{i=1}^{N_+} n_i \left(|M_{(+)i}|^2+|\widetilde M_{(+)i}|^2\right)\right.
\\&&\hspace{-20pt}-\left.
\frac{1}{2} \sum_{i=1}^{N_-} m_i \left(|M_{(-)i}|^2+|\widetilde M_{(-)i}|^2\right)+\frac{\xi^2}{T+\overline{T}}
\right]^2,
\end{eqnarray}
where $\xi^2=3 \delta_{GS}/2$ is related to the $\xi_{FI}$ introduced above as $\xi^2 = \xi_{FI}^2 (T+\overline{T})$,
and we considered for concreteness a
gauge kinetic $f_{U(1)_X} = (T/2 \pi)$  (more general gauge kinetic
functions $f_{U(1)_X} = f_0 + f_1 T $ do not change qualitatively our
analysis below).  
(ii) the superpotential of Eq.~(\ref{superpotential}), and (iii)
assuming for simplicity canonical  kinetic terms
for the $\Phi$ and messengers fields and the standard $K=-3 \log(T+\overline{T})$ K\"ahler potential
for the modulus $T$.
We observe that a more general K\"ahler potential for the $\Phi$ and messengers fields would not affect
sensibly the form of the potential around the relevant metastable minimum,  where these fields have
small, or zero, vev's.

The minimization of the complete potential is a complicated problem. We face it in two separate steps:
we first consider the minima where all the messengers have zero vev's. These minima are the
only ones we consider for phenomenology, since a vev for a messenger would induce an undesired breaking of
the SM gauge symmetry. In the second step we consider the metastability of these minima, by studying
the possible extrema with non-zero messenger vev's, and show that these extra vacua are ``far enough''
to ensure the metastable vacuum to be long lived. Notice that for the
minimal model \cite{dmpr}, the SUSY breaking vacuum was the absolute
ground state. The new instabilities we are discussing here appear due
to the new messengers $M_0$ and $M_-$. 

%%%%%%%%%%%%%%%%%%%%%%%%%%%%%%%%%%%%%%%%%%%%%%%%%%%%%%%%%

\subsubsection{The metastable vacuum}
\label{minimization}
Since the messenger fields cannot appear linearly in the potential, the point $M=0$, $\widetilde M=0$,
$\partial_{\Phi_+} V=0$, $\partial_{\Phi_-} V=0$, $\partial_{T} V=0$, is an extremum of the potential.
Thus, an effective approach to the minimization is to consider the locus  $M=0$, $\widetilde M=0$,
minimize the reduced potential w.r.t. the other fields, and then check the mass matrix for the messengers
to ensure that the extremum is actually a minimum.
The reduced potential has the form
\begin{eqnarray}
V&=&
\frac{e^{-2 bt}}{(2t)^3}\left[
\frac{1}{3} \left|a (3+2bt) \Phi_-^q (\Phi_- \Phi_+)^p+3 e^{b T} (W_0+m \Phi_- \Phi_+)\right|^2
\right.\\&&\left.
+\left|a \Phi_-^{q-1} (p+q+|\Phi_-|^2) (\Phi_- \Phi_+)^p+
e^{b T}(W_0 \bar\Phi_-+m \Phi_+ (1+|\Phi_-|^2) )\right|^2
\nonumber
\right.\\&&\left.\nonumber
+\left|a \Phi_-^{q+1}  (p+|\Phi_+|^2)(\Phi_- \Phi_+)^{p-1}+
e^{b T} (W_0 \bar\Phi_++m\Phi_- (1+|\Phi_+|^2))\right|^2
\right.\\&&\left.
-3 \left|e^{b T}\left(W_0+m \Phi_- \Phi_+\right)+a  \Phi_-^q (\Phi_- \Phi_+)^p\right|^2
\right]
\nonumber
+\frac{2 \pi}{t}  \left(\frac{\xi ^2}{2t}-|\Phi_-|^2+|\Phi_+|^2\right)^2
\end{eqnarray}
where we have neglected, in the prefactor $e^K$ the terms $e^{-|\Phi_-|^2-|\Phi_+|^2}$
since we expect the vev's of the $\Phi$ fields to be small, and we defined $2t=T+\bar T$.\\
The minimization can be done along the lines of \cite{dmpr}: in the $p=0$ case,
we have 
\begin{equation}
F_{\phi_+}\sim e^{{\cal K}/2} m \phi_- \ \gg F_{\phi_+}, \ F_{T} 
\end{equation}
with in particular the contribution of the modulus F-term
parametrically described by $\tilde \epsilon \ = \ \sqrt{\frac{F^T F_T}{F^{+}F_{+}}} = \frac{4+3q}{2 b t} \ $.
Therefore uplifting requires 
$F_{\phi_+} = \sqrt{3} m_{3/2}$ and thus
$W\sim W_0$ in the minimum, while 
\begin{equation}
\phi_-^2\sim \xi_{FI}^2 \sim \frac{3q}{2bt}\ ,\quad \phi_+\sim -\phi_-^2/\sqrt{3} \sim -\frac{\sqrt{3}q}{2bt}
\end{equation}
and $t$  being implicitly given by $bt e^{-b T}\sim  3W_0 /2a\phi_-^q$.
We can also link the value of the fields to some phenomenological quantities, such as
$bt = \log{\left( m_{3/2}^{-1}\right)}  + \kappa$ where $\kappa \sim {\cal O} (1)$.
Using the notation of Sect. 2, the SUSY breaking field $S$ must be
identified with $\phi_+$, and
\begin{equation}
\frac{F_{S}}{S}\equiv \frac{F_{\phi_+}}{\phi_+}\sim \frac{m_{3/2}}{S}
\end{equation}
with $S\sim \frac{\sqrt{3}q}{2bt} \ll 1$.
On the other hand
\begin{equation}
\frac{F_{\phi_-}}{\phi_-} \sim m_{3/2} \ .
\end{equation}

\begin{figure}[t]
\mbox{}\hspace{12pt}
\epsfig{file=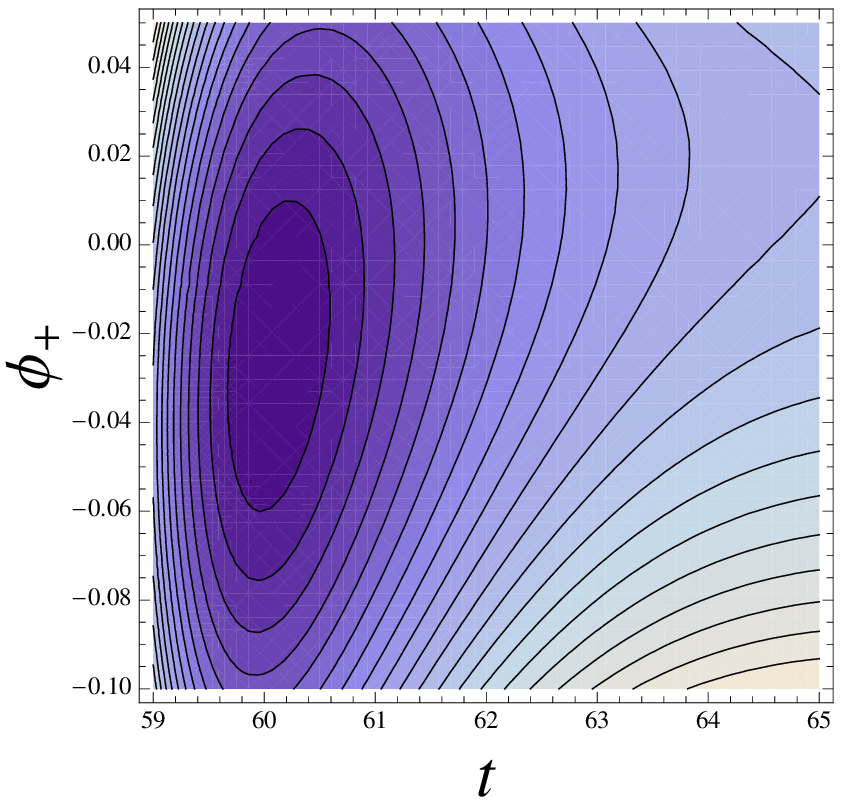,width=.4\textwidth}\hspace{24pt}
\epsfig{file=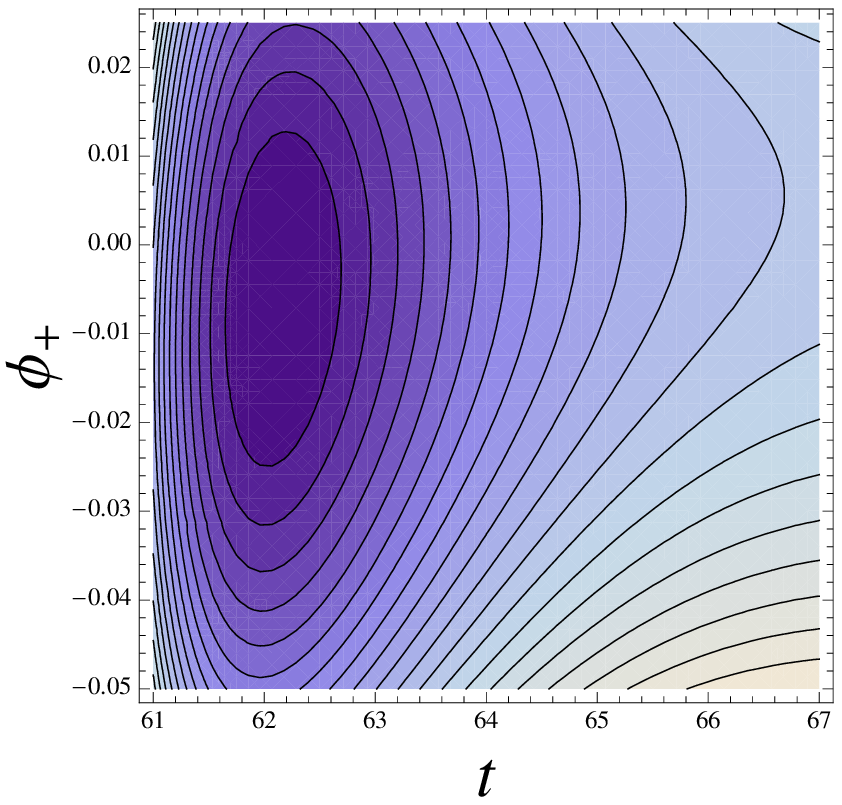,width=.4\textwidth}\vspace{-12pt}
\caption{\footnotesize
Minimum for the potential reduced as a function of $\phi_+$ and $t$ only, $\phi_-$ having
been stabilized at $~\sqrt{3q/2bt}$ by its e.o.m. (substantially dominated by the D-term potential).
The plot on the left is made for  $W_0\sim -4\times 10^{-13}$, $p=0$, $b=1/2$, $a=1$, $q=1$,
and shows the presence of an uplifted minimum (the uplifting is guaranteed by choosing
the parameter $m\sim \sqrt{3} |W_0|/|\phi_-|$) at $\phi_+ \sim -\phi_-^2/\sqrt{3}\sim 0.02$, $t\sim 60$
in $M_{Planck}=1$ units. The plot on the right is made changing $q=1\rightarrow q=1/4$, so that
the minimum value of $\phi_-$  is one quarter than previously, and $t$ is slightly increased.}
\label{fig:minp1}
\end{figure}

We also have a good numerical control over the minimum, since under
the assumption that $e^{-b T}$, $W_0$, $m$
$\ll 1$, the derivatives of the potential are dominated by the D-term contribution, and thus the field
$\phi_-$  is frozen by the requirement of a small D-term. In other words, we can consider the e.o.m.
for $\phi_-$, and study the potential on the locus  $\partial_{\phi_-}V=0$.
Such a reduced two-field problem can be easily approached numerically, finding good agreement with the
analytic results. In particular, on the left hand side of
Fig.~\ref{fig:minp1} we consider an uplifted minimum 
with $p=0$, $q=1$,
$b=1/2$, $a=1$, and $W_0\sim -4\times 10^{-13}$, so that $m_{3/2}\sim 4 \times 10^{-16}$ (all the
dimensionfull quantities are in Planck units);
in this case the analytic study would fix $\phi_-\sim 0.22$, $\phi_+\sim -0.029$, $t\sim 60$, in a very
good agreement with the numerical outcome. On the right hand side of Fig.~\ref{fig:minp1}, instead, we show
the $q=1/4$ case, keeping the other parameters at the same values. One can see that, as analytically
computed, the value of $\phi_+$ at the minimum is now roughly $1/4$ than before, and the value of
$t$ slightly increases.

\begin{figure}[t]
\begin{center}
\epsfig{file=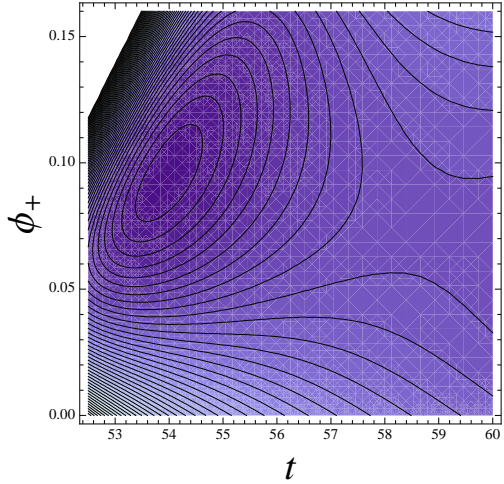,width=.4\textwidth}
\end{center}\vspace{-24pt}
\caption{\footnotesize
Minimum for the potential reduced as a function of $\phi_+$ and $t$ only, $\phi_-$ having
been stabilized by its e.o.m. (substantially dominated by the D-term potential).
The plot  is made for  $W_0\sim -4\times 10^{-13}$, $p=1$, $b=1/2$, $a=1$, $q=1$,
and shows the presence of an uplifted minimum (the uplifting is guaranteed by choosing
the parameter $m\sim \sqrt{3} W_0/\phi_-$) at $\phi_+ \sim 0.1$, $t\sim 54$
in $M_{Planck}=1$ units.}
\label{fig:minp0}
\end{figure}

If $p> 0$ an analytic study of the minimum is much harder, but we can still approach the problem
numerically, as explained above. We obtain, in the $p=1$ case, that the value of $\phi_-$ doubles due to
a dramatic growth of $\phi_+$, so that $\phi_+\sim \phi_-$,  while the value of $t$ slightly decreases
(see Fig.~\ref{fig:minp0} for the minimization in the case $W_0\sim -4\times 10^{-13}$, $b=1/2$, $a=1$, $q=1$
and $m\sim \sqrt{3} |W_0|/|\phi_-|$ to ensure the uplifting of the minimum).

The minima we found are consistent and (meta)stable provided that the messenger fields, that
we assumed to have zero vev, have positive masses. This can be always ensured, by properly
choosing the various messenger couplings. On the other hand, the absence of tachyonic modes
for the messengers is much weaker than the requirement that the messenger masses are
phenomenologically viable. Thus, we postpone the study of the
phenomenological bounds on the messenger couplings
to the following sections.

%%%%%%%%%%%%%%%%%%%%%%%%%%%%%%%%%%%%%%%%%%%%%%%%%%%%%%

\subsubsection{Lifetime of the metastable vacuum}

In this section we consider the metastability of the vacuum by
studying the  other possible
vacua that can be present in case we allow the messenger fields to
get non-zero vev's.
Our aim is to prove that the probability decay is very small, the latter being given, in the semiclassical
case and in the triangular approximation \cite{duncan}, by $e^{-S_b}$,
$S_b=(\Delta \Phi)^4/\Delta V$, with $\Delta \Phi$ the typical variation of the fields in
passing from a metastable minimum to a stable one, and $\Delta V$ being the corresponding variation
of the potential.
We approach this problem in a simplified version, assuming that only one messenger field develops
a vev, and we distinguish the three cases in which such a field  (i) is neutral (ii) has positive charge,
(iii) has negative charge; moreover, we consider here only the rigid SUSY case, neglecting the
$T$-modulus field.
This is reasonable since eventual minima in which the position of $T$ varies sensibly are harmless. Indeed,
the only problem would be a strong decrease of $T$, but in such a case we can estimate
$\Delta V <e^{-bt}/(2t)^3$, while $\Delta \Phi\sim\Delta t$. Thus, we have $S_b\sim 8 e^{bt} (t_0-t)^4 t^3 $
with $t_0\sim 10^2$, and, assuming $t\ll t_0$, $t>1$, we have $S_b\gg 1$.
In a similar way, we can argue that any vacuum displaced from the metastable one by a large change
in the other fields is harmless for metastability. Indeed, since all the fields (but $T$) enter polynomially
in the superpotential, and have canonical kinetic terms, we can argue that $\Delta V\sim W_0^2/(2t)^3$,
thus $S_b\sim (\Delta \Phi)^4 W_0^{-2} (2t)^3\sim  (\Delta \Phi)^4 10^{30}$ in our case.
Thus, in what follows we will consider as dangerous only minima where the field vev's are very close
to the metastable case.

\paragraph{Neutral Messengers}
\mbox{}\\\noindent
In this case we consider the fields $\phi_+$, $\phi_-$, $M_{(0)}$, $\widetilde M_{(0)}$ having canonical
K\"ahler potential.  For these fields we take the superpotential (in
the simple case $q=1$. The more general case $q \not=1$ is similar) 
\begin{eqnarray}
\label{simpsupchg0}
W=m\phi_+ \phi_- +(\lambda_{0} \phi_+\phi_- +\mu) M_{(0)}\widetilde
M_{(0)}+\Lambda \phi_- \ . 
\end{eqnarray}
That is a simplification of the general form given previously, in which we neglect the constant term $W_0$,
irrelevant in the rigid case, and we introduce the constant term $\Lambda$ to mimic the term $a e^{-bT}$.
Introducing  the D-term potential
\begin{equation}
V_D=\frac{g^2}{2} D^2,\,\,\,D=\phi_+^2-\phi_-^2+\xi_{FI}^2 \ , 
\end{equation}
we can compute the rigid potential. With this at hand we can first check the extremum
at zero messenger vev's, i.e. our metastable minimum. We find
\begin{equation}
\phi_-^2 \sim \xi_{FI}^2,\,\,\phi_+\sim-\frac{\Lambda}{2m},
\end{equation}
in qualitative agreement with the study described above including all the SUGRA corrections.
The obtained $\phi_+$ value is only an order of magnitude
estimate, but we will see that this is more than enough for our purposes.
We can now consider the extra solution in the presence of non-zero messenger vev,
looking for solutions ``close'' to the previous one.
In particular, assuming $\phi_+ \ll \phi_-$ and $\phi_{-}^2 \sim \xi_{FI}^2$, we find that the e.o.m.'s
admit a single solution, in the $\mu=0$ limit, for
\begin{equation}
\phi_-^2=\xi_{FI}^2,\,\,\phi_+=0,\,\,M_{(0)}\widetilde M_{(0)} =-
\frac{m}{\lambda_{0}} \ . \label{neutral01}
\end{equation}
We see that this solution is distinct from the previous one. Therefore, even assuming large
SUGRA corrections, we expect $\Delta\phi_+\sim \Lambda/m$, big enough to ensure
a long-lived metastable vacuum.
This result is stable against perturbations due to the presence of a non-zero
$\mu$, as long as $\mu$ remains small enough. For large $\mu$, on the
other hand, the minimum (\ref{neutral01}) disappear altogether. 

 \paragraph{Negative charge messengers}
 \mbox{}\\\noindent
In this case we consider the fields $\phi_+$, $\phi_-$, $M_{(-)}$, $\widetilde M_{(-)}$ having canonical
K\"ahler potential and superpotential
\begin{eqnarray}
\label{simpsupchgmin}
W=m\phi_+ \phi_- +\lambda_{+} \phi_+ M_{(-)}\widetilde M_{(-)}+\Lambda
\phi_- \ ,
\end{eqnarray}
motivated as above. We took for illustration messenger charges
$-1/2$~; again the result is qualitatively similar for other negative
charges. 
The F- and D-terms are
\begin{eqnarray}
&&D=|\phi_+|^2-|\phi_-|^2-\frac{1}{2}|M_{(-)}|^2-\frac{1}{2}|\widetilde M_{(-)}|^2 +\xi_{FI}^2 \ ,\,\,\,\,\,\,
\nonumber\\
&& F_{\phi_+}=m\phi_- + \lambda_{+} M_{(-)}\widetilde M_{(-)},\,\,\,\,\,\,
F_{\phi_-}=m\phi_+ +\Lambda,
\\\nonumber
&&F_{M_{(-)}}=\lambda_+ \phi_+ \widetilde M_{(-)},\,\,\,\,\,\,
F_{\widetilde M_{(-)}}=\lambda_+ \phi_+ M_{(-)}.
\end{eqnarray}
In case we neglect the $\Lambda$ corrections,
a SUSY minimum is present at $\phi_+=0$, with the vev's of $\phi_-$ and  $M_{(-)}$, $\widetilde M_{(-)}$
fixed by the $D=0$ and $F_{\phi_+}=0$ conditions.
The minimum is actually a one-dimensional flat direction in the $\phi_-$, $M_{(-)}$, $\widetilde M_{(-)}$
space, that can be dangerously close to the metastable minimum only in case the messengers
have similar vev's (if an hierarchy is present, then the value of $\phi_-$ is significantly
smaller than in the metastable vacuum, and thus we do not expect a fast decay of the latter in the new
minimum). We can thus study the most problematic case $M_{(-)}\sim \widetilde M_{(-)} = M$.
In this case, fixing $\phi_-$ via the F-term condition, we have the requirement
\begin{equation}
\frac{\lambda_+^2}{m^2} |M|^4+|M^2|=\xi_{FI}^2 ,
\end{equation}
that has solution $M^2\sim\xi_{FI}^2$ for $\lambda_+\ll m$, and $M^2\sim\xi_{FI} m/\lambda_+$ for
$\lambda_+\gg m$. The first (as well the ``intermediate'' case $\lambda_+  \sim m$) is harmless, since the new
vacuum is ``far'' from the metastable one. A possible problem may arise in case $\lambda_+\gg m$,
since the variation of the messenger fields vev in this case is very small. In such case,
we argue that the vev $\phi_+$  varies sensibly, such that the new
vacuum is ``far'' enough from the metastable one, that remains long lived.
In order to have an
estimate of $\phi_+$ in the new minimum, we have to introduce the $\Lambda$ corrections that, in the
metastable case, lead to $\phi_+\sim - \Lambda/2m$.
In the present case we can use the e.o.m. and observe that, in the approximation $\lambda_+\gg m$,
$\phi_+$ is stabilized at values much smaller than $\Lambda/2m$. In other words, $\phi_+$ is stabilized
due to a different mechanism then before.
Indeed, in the presence of a non-zero messenger vev, we observe a tension between the
$F_{\phi_-}=0$ and the $F_{M_{-}}=0$ conditions, not present in the metastable case, that pushes
$\phi_+$ to smaller values. Thus, we expect this solution for $\phi_+$ to be distinct from the metastable
solution, even when considering the whole SUGRA potential, so that
$\Delta\phi_+\sim \Lambda/m$, big enough to ensure the metastable minimum to be long lived.

\paragraph{Positive charge messengers}
 \mbox{}\\\noindent
 In this case it was checked in \cite{dmpr} that no dangerous new
 vacua are present. This provides an interesting example of a
 messenger sector that introduces no dangerous minimum breaking
 charge/color. On the other hand, that model does not
 realizes a standard GMSB transmission, but a combination of gravity
 and non-standard gauge contributions.    

%%%%%%%%%%%%%%%%%%%%%%%%%%%%%%%%%%%%%%%%%%%%%%%%%%%%%%%%%%%

\subsubsection{Constraints and assumptions for messengers} \label{messengers}

Important constraints have to be imposed on the messengers sector from
consistency requirements of the 
theory, but also from a phenomenological point of view. These
constraints concern in particular the number 
of the messengers, the relation between their relative charges and
their couplings to the supersymmetry breaking fields $\Phi_+$ and $\Phi_-$.\\
As discussed above, the only constraints on the numbers and charges of
messengers coming from the anomalies, is determined by the relation
\begin{equation} \label{rel-charg}
\Sigma_{i=1}^{N_+} n_i - \Sigma_{j=1}^{N_-} m_j = c  \ , 
\end{equation}
where $c$ is dictated by the peculiar expressions for the gauge
kinetic functions . Therefore, in particular, in the simplest case
where all the charges are 
$\left(\pm  q \right)$ and $c>0$, the relation (\ref{rel-charg}) is translated in $N_+ - N_- > 0$.\\
Nonetheless, it is well known \cite{gr} that whereas the
presence of messenger fields at an intermediate scale does not modify
the value of $M_{GUT}$, the inverse 
gauge coupling strength at the unification scale $M_{GUT}$ receives extra contributions
\begin{equation}
\delta \alpha_{GUT}^{-1} = - \frac{N}{2\pi} \log{\left( \frac{M_{GUT}}{m_f} \right)}\ ,
\end{equation}
with $N$ the number of messengers (when all of them are in the
fundamental representation of $SU(5)$) 
and $m_f$ their supersymmetric (fermionic) mass. Therefore there is an
upper bound on the total number 
of messengers given by the request of perturbativity of gauge
interaction up to the cutoff scale of 
the model, and depending on their mass scale. Thus a conservative choice is
\begin{equation} \label{Nmess}
(N_+ + N_- + N_0) \log{\left( \frac{M_{P}}{m_f} \right)}  < 150 \ , 
\end{equation}
in the approximation that all the messengers are at the same scale
$m_f$. Moreover, this is also a necessary 
condition in order to be able to do a phenomenological numerical
analysis. In fact, if some messengers 
appear at low-energy scale, and others at high-energy, the dynamics at
intermediate scale 
is governed not only by the MSSM fields, but also by the low-energy
messengers. This complicate the RG 
flow equations of the theory and makes the model difficult to study,
since the precise results really f
depend on the details of the model. Therefore, for simplicity, during
our analysis we assumed all the 
messengers having in first approximation the same high-energy
supersymmetric mass (of the order of the GUT scale), since in this way 
they do not change the RG flow equations for soft terms.\\
However it is useful to discuss in any case the limits on the
couplings $\lambda$. Referring to the 
minimization procedure in the case $p=0$, we can write the conditions
that couplings, or equivalently, 
the supersymmetric masses have to satisfy in order to keep the minimum
stable  and do not introduce tachyonic directions. This condition
comes from requiring positive 
eigenvalues for the mass scalar matrix (\ref{matrixmess}). \\
It is easy to see that for the messengers with positive $U(1)_X$
charges, the eigenvalues are 
automatically positive once reasonable charges are imposed.\\
The conditions for the negatively charged and neutral messengers read instead
\begin{eqnarray} \label{tach-mess}
(m_f)_j = \lambda_{(-)j} |\Phi_{+}|^{m_j}&>& 
\sqrt{3} m_j |\Phi_{+}|^{-1} \left( 1 + \frac{|\Phi_+|}{2 \sqrt{3} m_j} \right) m_{3/2}  \simeq 10^2 m_j m_{3/2} \\
(m_f)_r = \lambda_{(0)r} |\Phi_{+} \Phi_{-}|^{l_r} c_r &>& l_r \left(
  \left| \frac{F_+}{\Phi_+}+\frac{F_-}{\Phi_-}
  \right|\right) \simeq 10^2 l_r m_{3/2} \ . 
\end{eqnarray}
Therefore, the limits imposed by consistency are not stronger than those imposed by phenomenology,
which require messengers much heavier than the electroweak scale. \\
Finally, as introduced at the beginning of the section, we could in
principle add to the superpotential (\ref{superpotential}) terms of
the form $\left[ \lambda \left( \Phi_+ \Phi_- \right)^{k} -\mu \right]
\left(a_1 M_{-} \tilde{M}_{+} + a_2 M_{+} \tilde{M}_{-} \right)$, if
there exists two couples of messengers with opposite $U(1)_X$
charge. Once diagonalized the mass matrix for $M_{+}, M_{-},
\tilde{M}_{+}\tilde{M}_{-}$, it is possible to see that the picture is
similar to that of the neutral messengers, with the introduction of
effective parameters like $c_r$ taking into account sums and
differences of masses. However in this case the masses are naturally
of the same order, or in other words these effective parameters $c_r$
can be very small and generate tachyonic directions. The final result
should then depend on the details of the model. 
Therefore, in what follows we will forbid these terms for simplicity.

%%%%%%%%%%%%%%%%%%%%%%%%%%%%%%%%%%%%%%%%%%%%%%%%%%%%%%
%%%%%%%%%%%%%%%%%%   PHENOMENOLOGY   %%%%%%%%%%%%%%%%%
%%%%%%%%%%%%%%%%%%%%%%%%%%%%%%%%%%%%%%%%%%%%%%%%%%%%%%

\section{Phenomenological consequences}
\label{pheno}

In what follows we investigate the effects of supersymmetry breaking
in the observable sector, that we take to be the
Minimal Supersymmetric Standard Model (MSSM).
As introduced above, the model described by the superpotential
(\ref{superpotential}) modifies in a non trivial way the results already
 obtained in \cite{dmpr}, due to the presence of the messengers
with opposite and neutral charges.
However, the main feature of the hybrid model is preserved:
 gravity and gauge mediation give comparable contributions to the soft terms.
The non--universality in the scalar sector induced by gauge mediation,
and the negative non--standard contribution induced by D--terms
have severe consequences on the mass spectrum
and relic density constraints.

%%%%%%%%%%%%%%%%%%%%%%%%%%%%%%%%%%%%%%%

\subsection{General parameterization of Hybrid Models}

The different contribution from the three mediation mechanisms
 (standard gauge, non-standard gauge and
gravity, see the Appendix for details) can be parameterized
by the gravitino mass $m_{3/2}$, and two dimensionless
parameters $\alpha$ (defined in Eq.~(\ref{alpha})) and $\beta$ which measure the relative size of
standard and non-standard gauge mediation contributions with respect to $m_{3/2}$;
by introducing them, we disentangle the microscopic description of a
hybrid model, given in Sect.~\ref{micro}, from its
phenomenological study.
We can then write the breaking terms:

\begin{eqnarray}
M_a &=& m_{3/2} \left( \tilde{\epsilon} + g^2_a S_Q~ \alpha \right) \nonumber \\
m_i^2 &=& m_{0}^2 + m_{3/2}^2 C_i S_Q \left( - \beta + \frac{\alpha^2}{N} \right),
\label{general}
\end{eqnarray}

\noindent
where $S_Q$ is the Dynkin index of the messenger representation
(1/2 for the fundamental representation of $SU(N)$),
$g^2_a$ are the gauge couplings and
$C_i = \sum_{a} g^4_a C^a_i$,  $C^a_i$
being the Casimir of the MSSM scalar fields representations
(in our normalization the Casimir of the fundamental representation of
$SU(N)$ is $(N^2-1)/(2N)$,  that of  $U(1)_Y$ is simply $Y^2$). The
gravitational contribution for the 
scalar fields is indicated here as $m_{0}^2$, but it will
be taken equal to $m_{3/2}^2$ in our analysis. The extra parameter $\tilde{\epsilon}$ includes the effects
of gravity mediation for gauginos. Whereas the coefficient relating
$m_0^2$ and $m_{3/2}^2$ is 
of order 1, we will see in the following that $\tilde{\epsilon}$ is instead of order ${\cal O} (10^{-1})$.
$N$ is the ``effective'' number of messenger fields contributing to GMSB.
Unlike the classical GMSB at low energy, 
the mediation in the hybrid models occur around the GUT scale where the
gauge contributions to the gaugino masses $M_a$
(proportional to their gauge couplings $g_a$) are approximately universal.
 Thus, the non-universality only affects the
scalars masses. Concerning the trilinear couplings $A_{i=t,b,\tau}$,
there is no 1-loop messenger contribution to the susy-breaking
trilinear terms. However $A_i$ terms are generated in the leading log
approximation by the RG evolution and are proportional to gaugino masses.
At GUT scale we will make no assumption on the value of the trilinear
couplings. They will be considered as free parameters thorough the
phenomenological study.
The reader can find in the appendix the explicit expression
of the mass terms for each generation of squarks and sleptons.

%%%%%%%%%%%%%%%%%%%%%%%%%%%%%%%%%%%%%%%%%%%%%%%%%%%%%%%%%%%

\subsection{Link with microscopic models}

The parametrization of eq.~\ref{general}, and in particular the allowed
values for the parameters therein, is the main information with
phenomenological relevance we extract
from any microscopic construction. 

In the high-energy model defined 
by the superpotential (\ref{superpotential}), after the minimization procedure shown in the section (\ref{minimization}),
in the first approximation it is possible to express all these
low-energy parameters in terms of  $m_{3/2}$ and the messengers masses and charges.\\
In a simplified case, where all
the subleading contributions are neglected, all the 
messengers are taken at the same mass scale $m_{mess}$ and couplings
$\lambda_i$, all the charges $n_i, m_j, l_r = 1$, 
the mass parameter $\mu_r = 0$ and, as before, $p=0$, the
correspondence is
\begin{eqnarray}
\tilde{\epsilon} &\simeq& \frac{(4+3q)}{2} \left[ \log{\left(
      \frac{1}{m_{3/2}}\right)}  + \kappa \right]^{-1} \ , \nonumber \\
N &\simeq& N_{-} + N_{0} \ , \nonumber \\
\alpha & \simeq & 
\frac{\sqrt{3}N}{8 \pi^2} \frac{1}{\langle\phi_+\rangle} \ \simeq \
\frac{N}{4 \pi^2 ~q} \left[ \log{\left(
      \frac{1}{m_{3/2}}\right)}  + \kappa \right] \ , \nonumber \\
- \beta &\simeq&  \frac{3}{(8 \pi^2)^2} \frac{1}{\langle \phi_- \rangle^2} c \left[ 1 - \log{\left( \frac{\Lambda_{UV}^2}{m_{mess}^2}\right)}
\right] 
\nonumber \\
&\simeq& \frac{1}{32 \pi^4~q} \left[ \log{\left(
      \frac{1}{m_{3/2}}\right)}  + \kappa \right]  c
\left[ 1 - \log{\left( \frac{\Lambda_{UV}^2}{m_{mess}^2}\right)}
\right] \ ,
\end{eqnarray}
where anomaly cancellation fixes $c \simeq N_+-N_-$, $c$ having been defined
in eq.~\ref{cdef}, and the SUSY breaking field $S$ has been identified with $\phi_+$.
Moreover, $\beta$ can be positive of negative; in what follows we consider the
most interesting case $\beta>0$. Finally, $\kappa$ is an ${\cal O} (1)$ parameter.
The qualitative picture does not change much in the most general case.
 Nonetheless in this approximation we can estimate a reasonable range
of values for these parameters in our phenomenological analysis, imposing some
 constraint required by consistency, coming for example from the link between
the gravitino mass and $\alpha$ and $\beta$, or the constraints on the numbers
 of messengers as discussed in section (\ref{messengers}), depending on their
 mass. 

In the
 models of Sections 4,5 $\alpha$
can vary between 0 and ${\mathcal O}(100)$, $\beta$ can vary between 0
and ${\mathcal O}(10)$.
Thus we can describe/motivate a regime where gravity mediation is dominant,
one where standard gauge mediation is (marginally) dominant and an
intermediate regime where the model is truly  hybrid.

%%%%%%%%%%%%%%%%%%%%%%%%%%%%%%%%%%%%%%%%%%%%%%%%%%%%%%%%%%%

\subsection{Gravity vs. GMSB}
Gauge mediation is dominant either by enlarging
the number of messengers (indeed, $\alpha$ grows
linearly with N), or by choosing the microscopic parameters to
 fix the vev's of the fields with
relevant F-term to small values. 
Moreover, we will consider $\tilde{\epsilon}= 10^{-1}$ in all our
phenomenological discussion, in a qualitative agreement with  the formula given above. It is evident that for $\alpha \sim 0$, the exact value of
$\tilde{\epsilon}$ should be very important for the numerical results.

It is easy to see that in a large region of the parameter space the
model is really hybrid, in the 
sense explained in section \ref{motivations}. In that region, the
parameters $\alpha, \beta, N$ and $q$, 
are $\sim {\cal O}(1)$,  whereas $\epsilon \sim {\cal
  O}(10^{-1})$. 

However, it is compelling to check if it is possible to obtain pure gravity or pure standard gauge
mediation in some corner of the parameter space.\\
Actually pure gravity mediation, or something undistinguishable from
it, is obtained when the parameters $\alpha$ and 
$\beta$ are negligible. This happens for example when the total
number of messengers is small and in 
particular $N \sim 0$ . In particular, the example analyzed in
\cite{dmpr} flows into this scenario by lowering the number of the
positively charged messengers or increasing their mass.

More difficult instead is to obtain pure standard gauge mediation, taking into account the condition (\ref{Nmess})
and the minimization result $\langle \phi_+ \rangle = - \xi_{FI}^2/\sqrt{3} = \sqrt{3} q / (2 b t)$.
The requirements are then $\frac{\alpha^2}{N} \gg 1$ but also $\frac{\alpha^2}{N} \gg \beta^2$. 
The following example illustrates the limits we can reach. Take $q = 1/5$,
$\tilde{\epsilon} \sim \frac{1}{15}$ and $N = 15$ and obtain:
\begin{eqnarray} \label{extreme}
\langle\phi_+\rangle\sim \frac{1}{400}\ , \quad m_{3/2} \sim 100 ~{\rm GeV} \qquad \alpha \sim 75 \qquad \beta
\sim 1 \ . \label{numbers} 
\end{eqnarray}
We are allowed to choose $q$ at an apparently unnatural small value (1/5), 
crucial in order to lower the $\langle\phi_+\rangle$ value, interpreting the 
coupling $e^{-bT} \Phi^q$ in the superpotential as coming from the
condensation of a strongly coupled sector. 
In this case in fact, $q \sim \frac{N_f}{N_c}$ with $N_f$ and $N_c$
respectively the number of flavors and colors 
of such a nonperturbative theory \cite{bd, dmpr}. Lowering the value of the gravitino mass acts in the direction
of increasing $\alpha$. However, while $\langle \phi_+ \rangle$ (and then $\alpha$) depends logarithmically
on $m_{3/2}$, the soft terms decrease proportionally to the gravitino mass. Fixed all the other parameters, the
phenomenology impose therefore a lower limit for the value for $m_{3/2}$. The resulting soft terms at high energy are
\begin{equation}
M_a \gtrsim 1~{\rm TeV} \qquad m_i^2 \sim (1~{\rm TeV})^2
\end{equation}

Although  the gauge to gravity contribution to scalar mass squared is only 100:1,
the diagonal  entries (in flavor space)  in the scalar mass
matrix are enhanced by the RG evolution to low energy,  making in this way negligible the
non-universal terms dangerous for the FCNC problem \cite{lavorone}.\\
However, after a more detailed analysis, described in the rest of this Section, it is possible to see
that at low-energy, even if the FCNC problem is reasonably solved in the squarks sector, it is still
open in the sleptons one, as shown in Table 1.
Finally, another possibility could be to decrease the factor $c_r$
coming from the contribution of 
neutral messengers. For these fields we are in fact allowed to add an
explicit mass term $\mu_r$ with $\mu_r +  \lambda_{(0)r} (\Phi_+
\Phi_-)^{l_r} \ll 1$, in order to enforce by tuning a light neutral messenger mass. The net effect is the increasing of the 
gauge mediation contribution coming from this kind of messengers;
therefore $\alpha$ grows keeping fixed 
all the other parameters. However, one has to be careful to not
produce tachyonic directions in messengers directions
(\ref{tach-mess}), so this possibility need to be investigated in more
detail before concluding about its viability. \\
As last remark, whereas gauge mediation provides a natural solution to the FCNC
problem of supersymmetric theories, it has serious problems in generating
the right order of magnitudes for the $\mu$, $B \mu$ mass parameters
of MSSM. On the other hand, hybrid models with a moderate gravity contribution ($m_{3/2} \sim 10-100$ GeV)
can provide a viable solution of the $B-\mu$ problem \cite{hybrid, Giudice:1988yz}.  \\

\subsection{Predictions-general discussion}

The main prediction of the class of models with anomalous $U(1)_X$ symmetry considered
in this paper is the existence of a lower bound on the vevs of the fields relevant for supersymmetry
breaking of order of $10^{-3}$ in the Planck mass units.  This implies an upper
bound on the parameter $\alpha$ present in Eq.~(\ref{general}) for the soft superpartner masses. 
The bound on $\alpha$  also depends on the acceptable number of messengers consistently with
the gauge coupling unification. For necessarily heavy messengers N can be even around 20.
Both effects together give an upper bound on $\alpha$ of order of 100 and this has important
phenomenological consequences.

It is convenient to organize our phenomenological discussion into two cases, depending on
weather the LSP is neutralino or gravitino. Indeed, a quick look at Eq. (\ref{general})
tells us that for $\alpha$  larger than $\sim 8$,
$m_{3/2}\lesssim 2 M_1 \mathrm{(GUT)}$, implying
$m_{3/2}\lesssim M_1 \mathrm{(EW)}$: for such values of
$\alpha$, the gravitino is the LSP and the stable relic candidate.

Thus, in the region where neutralino is the LSP (small values of $\alpha$),  gravity mediation
contribution to the soft masses remains important or can even be the dominant one. 
The sfermion soft masses are then  necessarily of the order of the graviton mass and, for instance,
the indirect limit on the stop mass following from the experimental lower bound
on the higgs mass implies gravitinos in the several hundreds GeV range.
However,
the new element of models with anomalous $U(1)_X$ and messengers is the presence of non-standard
gauge mediation contribution to sfermion masses generated by D-terms.
As already shown in \cite{dmpr}, when gravity and gauge mediation are comparable, they lead to
interesting effects in the sfermion spectra.
As will be shown below in a couple of concrete examples, the predictions of the model in this 
parameter range can be fully consistent with all experimental constraints  and in particular
neutralino is a good dark matter candidate.

For larger values of $\alpha$ gravitino becomes the LSP  and the role of gauge mediation is increasing. 
However, the first important point to notice
is that, since there is an upper limit on $\alpha$, the  ratio  of soft masses to the gravitino mass
remains bounded from above and the mentioned above limits on the stop mass imply a lower
limit on the gravitino mass of order 50-100 GeV  (the role of the left- right mixing in the stop sector, i.e. of A 
terms is important here). This is interesting since gravitino LSP in such  mass
range allows for high reheating temperature, consistent with leptogenesis.

However, important constraints on such scenarios come from the life time of the NLSP, to be consistent with
nucleosynthesis. It is know that neutralinos as NLSP are acceptable only if very
heavy.  Better candidates for NLSP are staus (for a recent discussion see \cite{steffen} and references
therein\footnote{We thank David Cerde\~{n}o for pointing out this issue to us.}), although
the lower bound on their masses is in a TeV range, too.
It is clear from Eq.~(\ref{general}) that  staus can be NLSP  only for sufficiently large N. 
The question of course is if such values are still consistent with our upper bound for $\alpha$.
A detailed study  of the full range of values of $\alpha$ is beyond the scope of this paper.
However,  as we  show below in our third example, the model can account for gravitino LSP  with around
100 GeV mass and a stau as NLSP with its mass in the TeV range.
To our knowledge, this is a rare example of a microscopic model  with such properties.

\subsection{Examples}

We illustrate the previous qualitative discussion  and the phenomenological viability of the
model in a few concrete examples.

The soft terms are defined in our parametrization (\ref{general}) by three free parameters at the GUT scale:
the gravitino mass $m_{3/2}$, the  standard gauge to gravity mediation ratio parametrized by $\alpha$
and the non--standard gauge to gravity mediation ratio parametrized
by $\beta$. Moreover, we have to take into account N, the number
of messengers, the $\mu$ mass-term and the bilinear $B$-term.
The absolute value of $\mu$ is determined by the minimization condition 
of the Higgs potential (assuming CP conservation), but its sign is not fixed.
Furthermore, instead of  $B$ it is more convenient to use the low energy parameter
 $\tan \beta = \la H_2^0\ra/\la H_1^0\ra$, which is a function of $B$ and
the other parameters.
Thus  the parameter space for a complete phenomenological study of
such  a model can be restricted to the following set of five parameters:

\begin{equation}
\label{e.params}
m_{3/2} \;, \ \hspace{.2cm} \   \alpha \;, \ \hspace{.2cm} \
\beta \;, \ \hspace{.2cm} \  N \; , \ \hspace{.2cm} \  \tan \beta \; , \ \hspace{.2cm}
\  {\rm sgn}(\mu) \; .
\end{equation}

In the following we first discuss the constraints we place upon the model.
It includes theoretical constraints (electroweak symmetry breaking
condition, color and charge breaking minima, dark matter abundance)
as well as current bounds from accelerator experiments. Then we present
our results.

\vskip .3cm
\noindent
\textit{(i) The mass spectrum constraints:}

\noindent
We have implemented in our analysis the lower bounds on the masses of
 SUSY particles and of the lightest Higgs boson.
In the squark and slepton sector we checked for the occurrence of tachyons.
We applied in our analysis the LEP2 lower bound limit on the mass of
the lightest chargino  $m_{\tilde{\chi}^+_1} > 103.5$ GeV.
In the non-tachyonic region, typically, the most constraining is the
 lightest Higgs boson mass constraint.
In the decoupling limit ($M_A \gg M_Z$, applicable in all our parameter
space), $m_h > 113.5$ GeV at $3\sigma$.
This bound is very sensitive to the value of the top mass. We have taken
$m_t=171$ GeV throughout our analysis.

\vskip .3cm
\noindent
\textit{(ii) The $b \rightarrow s \gamma$ branching ratio:}

\noindent
One observable where SUSY particle contributions might be large is the
radiative flavor changing decay $b \rightarrow s \gamma$. In the
Standard Model this decay is mediated by loops containing the charge
 2/3 quarks and $W-$bosons. In SUSY theories additional contributions
come from loops involving charginos and stops, or top quarks and charged
higgs bosons. The measurements of the inclusive decay
$B \rightarrow X_s \gamma$ at CLEO \cite{cleo} and BELLE \cite{belle},
 leads to restrictive bounds on the branching ratio $b\to s\gamma$.
 We impose in our analysis
$2.33\times 10^{-4}\leq BR(b\to s\gamma)\leq 4.15\times 10^{-4}$
 at the 3$\sigma$ level. Our analysis considers standard model
contributions up to the next to leading order (NLO) and takes into
consideration the tan$\beta$ enhanced contributions in MSSM beyond
the leading order. We mostly choose $\mu >0$ allowing cancelations
between chargino and charged Higgs contributions.

\vskip .3cm
\noindent
\textit{(iii) The anomalous moment of the muon:}

\noindent
We have also taken into account the SUSY contributions to
the anomalous magnetic moment of the muon,
$a_{\mu}= (g_\mu -2 )/2$.
We used in our analysis the recent experimental results for the muon
anomalous magnetic moment \cite{g-2}, as well as the most recent
theoretical evaluations of the Standard Model contributions
\cite{newg2}.
 It is found that when $e^+e^-$ data
are used the experimental excess in $(g_\mu-2)$ constrains a
possible SUSY contribution to be
$7.1 \times 10^{-10}\lesssim a^{\rm SUSY}_\mu \lesssim 47.1 \times 10^{-10}$ at
$2\sigma$ level. However when tau data is used a smaller
discrepancy with the experimental
measurement is found.
In order not to exclude the latter possibility, when analyzing
the parameter space with $\mu > 0$  we will simply plot contours
with the relevant value $a^{\rm SUSY}_\mu= 7.1\times 10^{-10}$.

\vskip .3cm
\noindent
\textit{(iv) Relic Density:}

\noindent
Our basic  assumption is  that the LSP is stable on cosmological time
scales. Furthermore we will assume that the LSP abundance is thermal.
Within such a framework, the regions of the parameter space that lead to
overproduction of dark matter are excluded. On the other hand, the regions
that yield LSP abundance below the WMAP limit are not considered as excluded
(though as less favored), but simply  require non--thermal production
or a dark matter candidate beyond the soft spectrum.
The WMAP collaboration gives the 3$\sigma$ narrow limit \cite{wmap}
\begin{equation}
0.087\lsim\relic h^2\lsim 0.138
\label{eq:WMAP}
\end{equation}
on the dark matter relic abundance.

\vskip 0.3cm
We show in table \ref{tab1} three examples of low-energy spectrum
in the case of ``standard'' hybrid mediation ({\bf A}),
with a non--standard contribution ({\bf B}) and
case  ({\bf C}) of a moderate dominance of  gauge mediation based on the example
of Eq. (\ref{extreme}).   The selected points  ({\bf A})  and   ({\bf B}) 
respect
accelerators and WMAP constraint. The example  ({\bf C})  is also consistent with accelerator
constraints and the gravitino LSP with 100 GeV mass is a good dark matter candidate for the reheating temperature
of  order $ 10^{-8}$ (see \cite{steffen} and references therein).

The points ({\bf A})  and ({\bf B}) are  selected from a big sample of points obtained
by the following procedure:
 Once  $\tan \beta$ and $sgn(\mu)$ are fixed (positive thorough our study),
 we scan over the gravitino mass
$0 < m_{3/2} < 2 $ TeV and $0 < \alpha < 10$.
The low energy mass spectrum  is calculated using the Fortran package
{\tt SUSPECT} \cite{Suspect} and  its routines  were described in detail
in ref. \cite{Suspect2}.
The evaluation of the $b \rightarrow s \gamma$ branching ratio,
 the anomalous moment of the muon and the  relic neutralino density
is carried out using the routines provided by the program
{\tt micrOMEGAs} \cite{micromegas}

The first remark is that the scalar particles are relatively heavier
than the gauginos. This mainly comes from the fact that scalars
receive at tree level a gravity-mediated contribution proportional
to $m_{3/2}$, whereas this contribution is suppressed
 by a factor $\tilde \epsilon$ for the gauginos, (Eq.\ref{general}).

\begin{figure}
    \begin{center}
\centerline{
       \epsfig{file=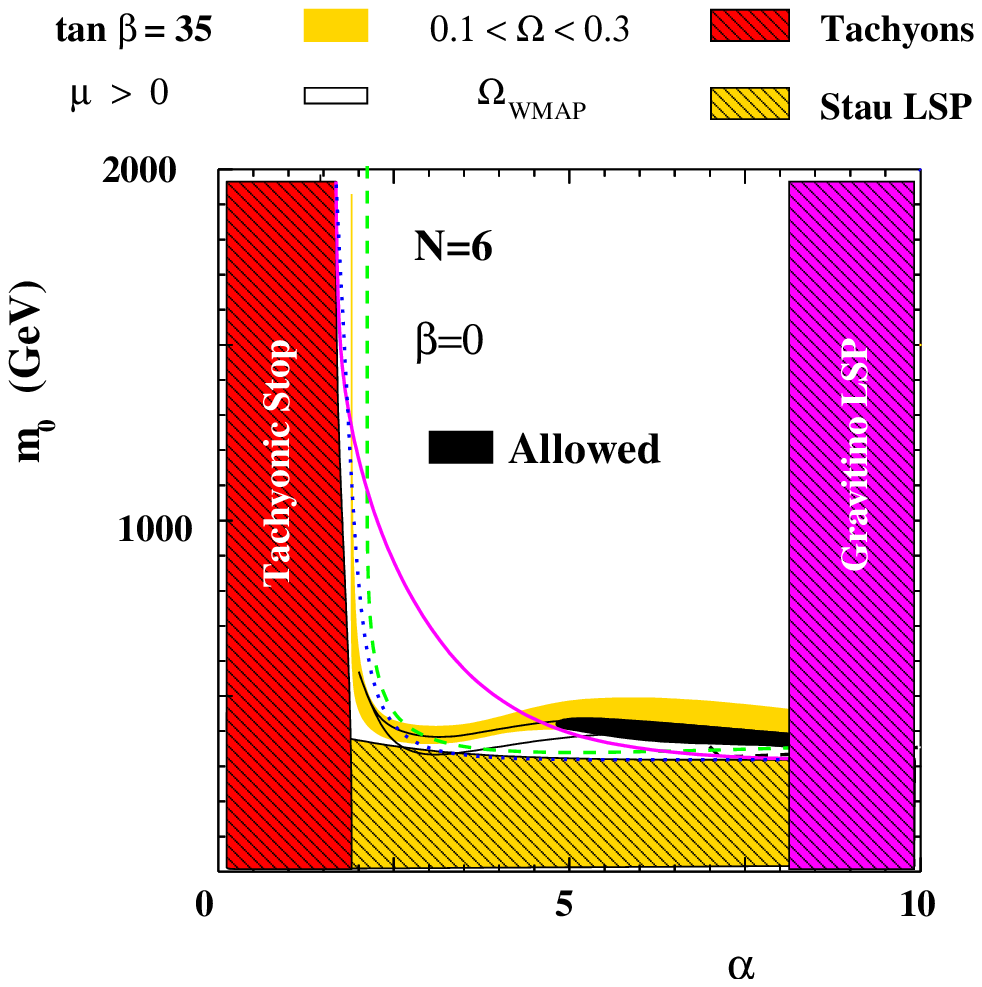,width=0.5\textwidth}
\hskip 1cm
       \epsfig{file=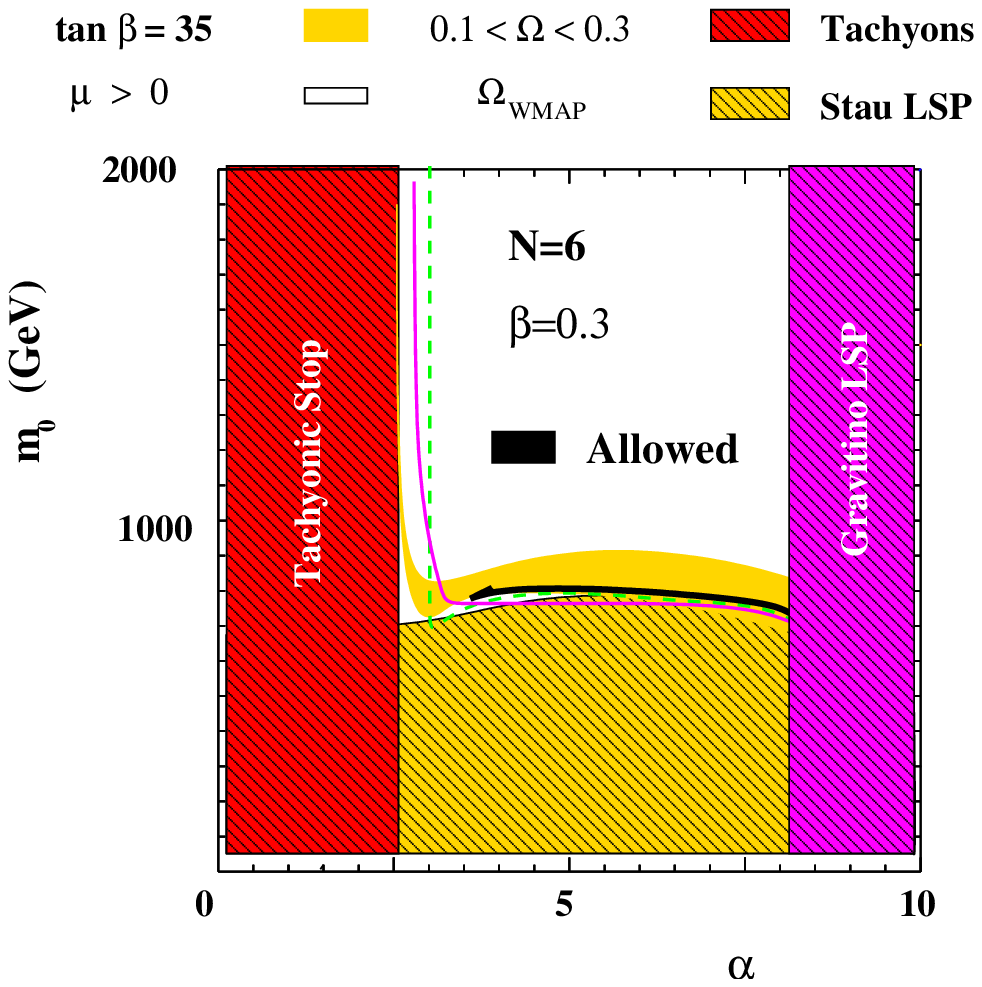,width=0.5\textwidth}
     }
          \caption{{\footnotesize
Scan on $m_0 versus \alpha$ with $A_i=-3 m_{0}$ and $\tan\beta=35$,
with 6 messengers and different values of $\beta$.
}}
\label{fig:A}
\end{center}
\end{figure}

\begin{figure}
    \begin{center}
\centerline{
       \epsfig{file=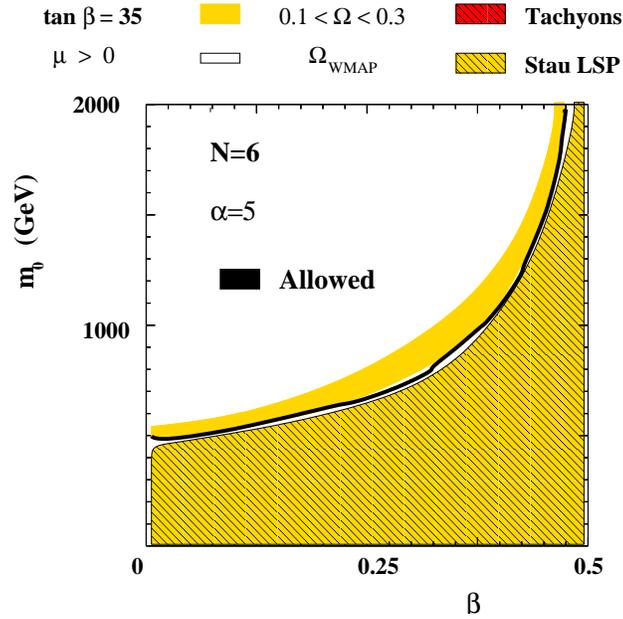,width=0.5\textwidth}
     }
          \caption{{\footnotesize
Scan on $m_0 versus \beta$ for $\alpha=5$, $A_i=-3 m_{0}$, $\tan\beta=35$,
and 6 messengers.
}}
\label{fig:beta}
\end{center}
\end{figure}

Concerning the influence of the negative non--standard gauge-mediated
term, if we compare points ({\bf A}) and ({\bf B}) we observe that
to obtain a similar amount of relic density, the model ({\bf B}) requires
a heavier gravitino. This comes from the fact that the WMAP
constraint is achieved through the coannihilation channel
$\chi^0_1-\tilde \tau_1$ and gauge-mediated
negative contributions acts on $m_{\tilde \tau_L}$ at GUT scale
 reducing considerably the $\tilde \tau_1$ mass at electroweak scale
(whereas not acting on the gaugino mass): we need a higher
value of $m_0$ to obtain $m_{\tilde \tau_1} \sim \chi^0_1$
where the coannihilation is efficient.
This effect is clearly depicted in Fig. \ref{fig:A} where the
parameter space excluded because the lightest stau is the LSP, increases
for increasing values of $\beta$. The cosmological allowed region
follows the line $m_{\tilde \tau_1} \sim \chi^0_1$ where the
coannihilation channel is dominant. We illustrate also the effect
in Fig. \ref{fig:beta} where we plot the allowed region as function of
$\beta$ : the parameter space is almost completely excluded for
$\beta \gsim 0.5$ because the $\tilde \tau_1$ becomes the LSP for any $m_{3/2}\lesssim 3$ TeV.

We also have calculated the mass spectrum for the case of moderate dominance
of gauge-mediation (point ({\bf C}) in table
\ref{tab1}). In this case, the 100 GeV gravitino is the LSP.
A relatively heavy  gravitino is necessary to respect the LEPII limit
on the higgs mass. The NLSP is stau and its mass is at the border line
of the limits given in \cite{steffen}. As mentioned earlier, no systematic study
has been performed of the parameter range corresponding to gravitino LSP.

\begin{figure}
    \begin{center}
\centerline{
       \epsfig{file=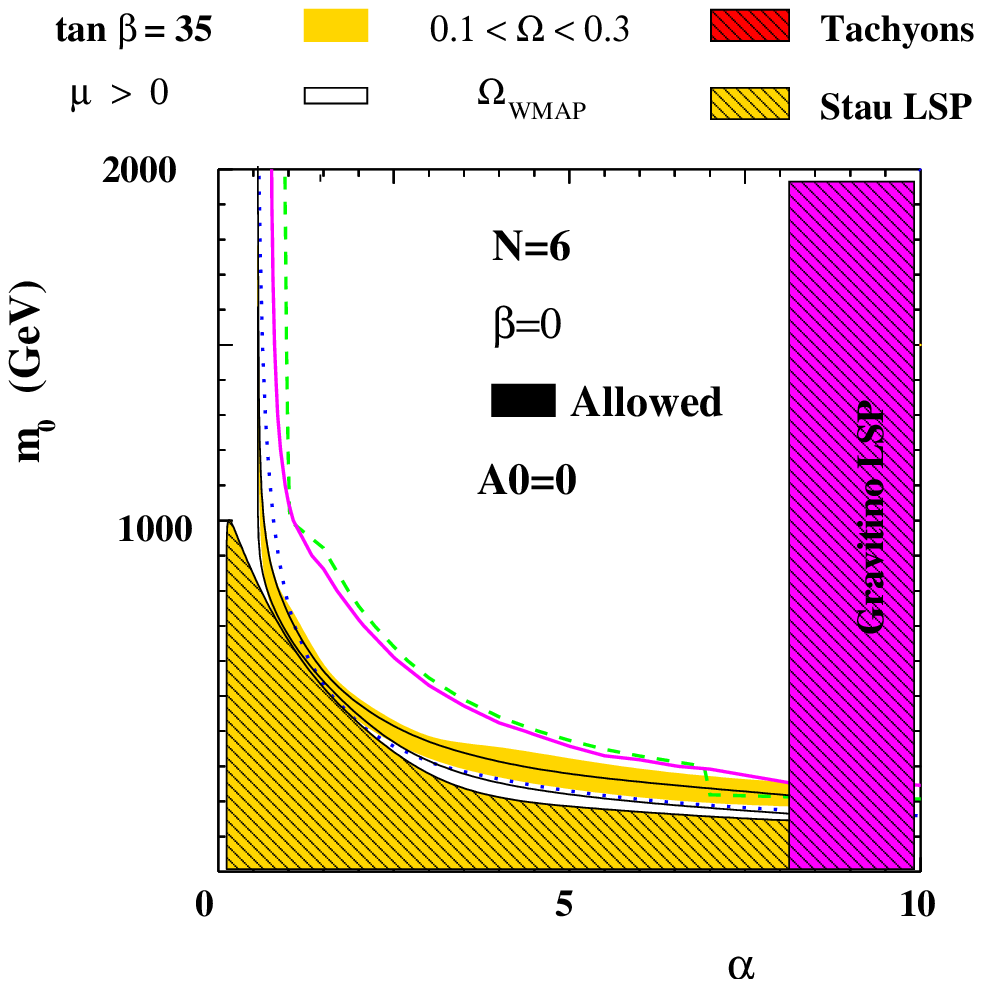,width=0.5\textwidth}
\hskip 1cm
       \epsfig{file=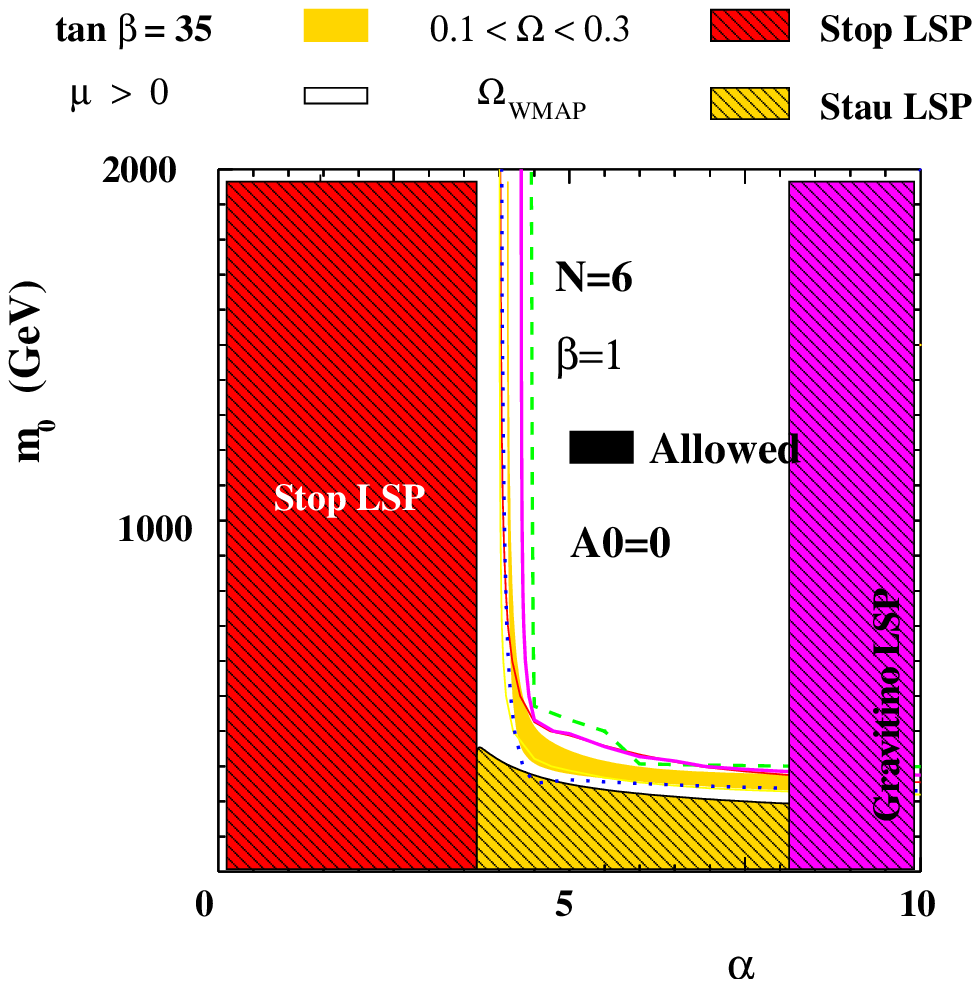,width=0.5\textwidth}
     }
          \caption{{\footnotesize
Same than Fig.\ref{fig:A} for $A_i=0$.}}
\label{fig:A0}
\end{center}
\end{figure}

We illustrate the influence of the trilinear coupling on
the parameter space  in Fig. \ref{fig:A0}, where
we reproduce the Fig. \ref{fig:A} except for $A_0=0$ at GUT scale.
We do not observe any point where the stop is tachyonic at the electroweak
scale, but a region where the stop can be the LSP for $\beta=1$.
However, no region of the parameter space respect the LEPII constraint
on $m_h$: lower values of $A_0$ implies lower radiative corrections
to the Higgs mass.

\begin{center}
\begin{table}
\centering
\begin{tabular}{|c|c|c|c|}
\hline
& \bf{A} & \bf{B}  & \bf{C} \\
\hline
$\mathrm{m_{0}}$ &385  & 1050 & 100  \\
$\mathrm{\alpha}$ &5   & 4 & 75  \\
$\mathrm{\beta}$ & 0 &  0.4  &  1  \\
N & 6 & 6 & 15   \\
$\tan \beta$ & 35 & 35 & 35 \\
\hline
$\mu$ & 890 & 1971 & 2070  \\
$M_1$ & 220 & 506 & 840  \\
\hline
$m_{\chi^0_1}$ & 218 & 504 & 830 \\
$m_{\chi^+_1}$ & 418 & 953 & 1542 \\
$m_{\tilde g}$ & 1207 & 2524 & 3943\\
\hline
$m_{h}$ & 118 & 124 & 122   \\
$m_{A}$ & 792 & 1716 & 1996  \\
\hline
$m_{\tilde u_1}$ & 1213 & 2500 & 3484  \\
$m_{\tilde t_1}$ & 747 & 1450 & 2852 \\
\hline
$m_{\tilde e_1}$ & 482 & 1303 & 865   \\
$m_{\tilde \tau_1}$ & 227 & 504 & 686  \\
\hline
$\Omega \mathrm{h^2}$ & 0.091 & 0.096 & * \\
\hline
\end{tabular}
\caption{Sample spectra. All masses are in GeV and $A_0=-3 m_0$} \label{tab1}
\end{table}
\end{center}

%%%%%%%%%%%%%%%%%%%%%%%%%%%%%%%%%%%%%%%%%%%%%%%%%%%%%%%%%%%%%%%%%%%%%%%%%%%%%

%%%%%%%%%%%%%%%%%%%%%%%%%%%%%%%%%%%%%%%%%%%%%%%%%%%%%%%
%%%%%%%%%%%%%%%%%%

\section{Conclusions and perspectives}

Supersymmetry breaking in models with gauged $U(1)$ symmetries and
Green-Schwarz mechanism is naturally realized in string theory
\cite{bd,Dudas:2005vv,dmpr,cw,hv}. Whereas at first sight, in this
framework gravity mediation, gauge mediation or a mixture of the two
are equally possible, in this paper we found that the constraints coming
from gauge invariance and moduli
stabilization are surprisingly strong. In particular we found that
explicit realizations of models in which gravity transmission could be
highly suppressed compared to the gauge contribution are very
difficult to obtain;  new supersymmetric vacua within the
supergravity regime appear and/or gauge mediation vacua are
incompatible with a small value of the cosmological constant. Our results therefore imply that models
with strong gauge dominance should probably contain at least two moduli
fields charged under the gauge
anomalous $U(1)$ symmetries, with a  highly-nontrivial dynamics. It is
useful to compare the situation to models with no gauged $U(1)$, where
the dynamics is much simpler and all possibilities of supersymmetry
breaking transmission are realized. From this viewpoint, the situation
is similar to the models of moduli stabilization and uplift of the
vacuum energy: whereas models with anomalous $U(1)$'s, very natural
from string theory perspective, are hardly compatible with a TeV
supersymmetry spectrum \cite{dm}, uplifts with F-terms are naturally compatible
with it \cite{F-term,dmpr}. On the other hand, models with purely F-term dynamics, with no
gauged $U(1)$ constraints of the type we discussed in the present
paper, are difficult to realize in string theory.

We generalized a previously proposed model which incorporated all the
constraints of gauge invariance and moduli stabilization by
considering the most general messenger sector compatible with anomaly
cancelation. The resulting model, coupled to MSSM,  automatically
contains gravity, standard and non-standard gauge contributions, which
are roughly  of the same order. In some regions of the parameter
space, standard gauge contributions can  moderately dominate over
gravity contributions and correspondingly the FCNC effects are below
per-cent level. It is however impossible without severe fine-tunings
to suppress further the gravity contributions.

The class of models we consider here is phenomenologically fully viable.
They can give either neutralino or gravitino as good dark matter
candidate. In the latter case, gravitino is necessarily relatively heavy (50-100 GeV).
This is cosmologically interesting since it is consistent with high reheating temperature needed
for leptogenesis.  The BBN constraints on the NLSP can be satisfied  (at least at a qualitative level
studied in this paper)  by the stau, for  number of messengers of order 15.

%%%%%%%%%%%%%%%%%%%%%%%%%%%%%%%%%%%%%%%%%%%%%%%%%%%%%%%%%%%
\section*{Acknowledgments}{ Work partially supported by the CNRS PICS \#~2530
and 3059, RTN contracts MRTN-CT-2004-005104 and MRTN-CT-2004-503369 and the
European contract MTKD-CT-2005-029466. E.D. and Y.M. would like to thanks the
Institute for Theoretical Physics of Warsaw form warm hospitality and
financial support via the ``Marie Curie Host Fellowship for Transfer of
Knowledge'' MTKD-CT-2005-029466. Y.M. would like to thank the European Network of
Theoretical Astroparticle Physics ILIAS/N6 under contract number
RII3-CT-2004-506222 and the French ANR project PHYS@COLCOS for
financial support. The work of A.R. and M.T. was supported by the European Commission Marie Curie Intra-European
Fellowships under the contract N 041443 and N 041273. The authors thank D. Cerde\~{n}o,
M. Cvetic, M. Lennek and G. Shiu
for useful discussions. E.D., S.P., A.R. and M.T. thank the Theory Group of CERN
for hospitality during the completion of this work.}

%%%%%%%%%%%%%%%%%%%%%%%%%%%%%%%%%%%%%%%%%%%%%%%%%%%%%%%%%%%

\appendix

\section{Appendix}

\subsection{Gravity mediation contribution}

The scalar mass matrix is defined as
\begin{equation}
M_0^2 = \begin{array}{|cc|}
m^2_{I \bar{J}} & m^2_{I J} \ , \\
m^2_{\bar{I} \bar{J}} & m^2_{\bar{I} J} \end{array} ~,
\end{equation}
where the various entries are  :
\begin{eqnarray}
m^2_{I \bar{J}} =~ \langle \partial_I \partial_{\bar{J}} V\rangle ~ =
~\langle  \nabla_{I} \nabla_{\bar{J}} V \rangle  \ , \\
m^2_{I J} =~ \langle \partial_I \partial_J V\rangle ~ =~ \langle  \nabla_{I} \nabla_{J} V \rangle \ .
\end{eqnarray}
The general expressions of the masses, for vanishing vacuum energy,
are of the form \cite{Dudas:2005vv}:
\begin{eqnarray}
\label{genscalar}
m^2_{I \bar{J}} &=& e^G ( G_{I \bar{J}} + \nabla_{I} G^{\bar{K}}
\nabla_{\bar{J}} G_{\bar{K}} - R_{I\bar{J}K\bar{L}} G^K
G^{\bar{L}} ~)
+ {1 \over 2} \sum_A g_A^2 D_A^2 ( G_{\bar{J}} G_I - G_{I \bar{J}} ) \nonumber \\
&-& \sum_A g_A^2 D_A (G_{\bar{J}} \partial_I D_A  + G_I \partial_{\bar{J}} D_A -
\partial_I \partial_{\bar{J}} D_A )
+  \sum_A g_A^2  \partial_I D_A \partial_{\bar{J}} D_A  \ , \\
\label{genbilinear} m^2_{IJ} &=& e^G(2  \nabla_J G_I + G^K \nabla_I
\nabla_J G_K ) - \sum_A g_A^2 D_A ( G_J \partial_I D_A + G_I
\partial_J D_A -\partial_I \partial_J D_A  ) \nonumber \\  &-& {1
\over 2} \sum_A g_A^2 D_A^2 ( G_I G_J + \nabla_I G_J) + \sum_A g_A^2
\partial_I D_A \partial_J D_A \ ,
\end{eqnarray}
where the function $R_{I\bar{J}K\bar{L}}$ is the Riemann curvature
of the K\"ahler manifold.

The standard results for the soft terms coming from the gravitational
effects, depend on some details concerning the coupling of the modulus
to the gauge multiplet of the 
MSSM, and the K\"ahler potential of the MSSM scalar fields. \\
From the point of view of a IIB string theory realization, irrespective on which type of brane MSSM sit (D7 or D3
branes), if they contain magnetic fluxes \cite{intersecting} the gauge kinetic functions
contain a T-dependence
\begin{equation}
f_a \ = \ {c_a \over 4 \pi} T \ + \ f_a^{(0)} \ , \label{sbt1}
\end{equation}
where $c_a$ are positive or negative numbers,  and $f_a^{(0)}$ effective
constants generated by the couplings of the MSSM branes to other,
stabilized
fluxes (e.g. the dilaton $S$). \\
Moreover, by denoting in what follows by $i,j$ matter fields and by
greek indices $\alpha$ any field contributing to SUSY breaking, a
relevant quantity for computing the soft terms is the coupling of
the matter fields metric $K_{i {\bar j}}$ to the SUSY breaking
fields. For our model in Sections \ref{micro} and \ref{pheno}, this can in turn be parameterized as
\begin{eqnarray}
&& K_{i{\bar j}}=(T+\overline{T})^{n_i} \left[ \delta_{i{\bar j}} +
(T + \overline{T})^{m_{ij}}|\phi_+|^2 Z'_{i\overline{j}} + (T +
\overline{T})^{p_{ij}}|\phi_-|^2 Z''_{i \overline{j}}
\right. \nonumber \\
&& \left. + (T + \overline{T})^{l_{ij}}(\phi_+ \phi_-
  Z'''_{i\overline{j}} + \mathrm{h.c})
+ O(|\phi_i|^4) \right], \label{sbt2}
\end{eqnarray}

\noindent where $G=K+\log|W|^2$, $K_{i{\bar j}}=\partial_i
\partial_{\bar j} K$, $i$ and $j$ representing the matter fields,
not participating to the SUSY breaking mechanism ($G_i=0$).  The
metric $K_{i {\bar j}}$ in (\ref{sbt2}) is written as an expansion
in powers of the charged vev fields $\phi_{\pm}/M_P \ll 1$, up to the
quadratic order.

%%%%%%%%%%%%%%%%%%%%%%%%%%%%%%%%%

\paragraph{Gaugino masses:}
The gaugino masses for a general gauge kinetic function $f_a$ are given
by \cite{bim}
\begin{equation}
M_a^{\mathrm{Grav.}}= \frac{\partial_T f_a}{Re[f_a]} e^{K/2} K^{T
\overline{T}} D_T W \ .
\end{equation}
\noindent
 For the phenomenological analysis, we use the hypothesis of a gauge kinetic function given in
(\ref{sbt1}) and in particular the unified case
\begin{equation}
\alpha_a = \ {c_a \over c_a + 4 \pi f_a^{(0)} / T} \ \simeq 1 \ .
\end{equation}
Under this assumption, using the definitions and the formulas given
in the previous sections we obtain
\begin{equation}
M_a^{\mathrm{Grav.}} = m_{3/2} \alpha_a \frac{(T+\overline{T})}{3} \
\frac{D_T W}{W}
 = m_{3/2} \alpha_a \frac{(T+\overline{T})}{3} \ G_T \  \simeq
 \tilde{\epsilon} ~m_{3/2} \ . \label{gmasses}
\end{equation}

%%%%%%%%%%%%%%%%%%%%%%%%%%%%%%%%%%%

\paragraph{Scalar masses:}

Using the classical formulas at the linear order in the D-term
\cite{Dudas:2005vv,bim}
\begin{equation}
{\tilde m}_0^2|_{i{\bar j}} \ = \ m_{3/2}^2 \left[ G_{i{\bar j}} -
G^{\alpha} G^{\overline{\beta}}
 R_{i \overline{j}\alpha \overline{\beta}} \right] \ + \ \sum_a g_a^2 D_a
 \partial_i \partial_{\bar j} D_a  \ ,
\end{equation}

\noindent and with the standard definitions

\begin{equation}
R_{i \overline{j} \alpha \overline{\beta}} \ = \
\partial_i \partial_{\overline{j}} G_{\alpha \overline{\beta}}
-\Gamma_{i \alpha}^m G_{m \overline{n}}
\Gamma^{\overline{n}}_{\overline{j}\overline{\beta}} \quad , \quad
\Gamma_{i \alpha}^m \ = \ G^{m \overline{k}}\partial_{\alpha}G_{i
  \overline{k}} \ ,
\end{equation}
for the neutral scalar mass terms we obtain, after normalization
of the kinetic terms :

\begin{eqnarray}
&& ({\tilde m}_0^2)_{i \overline{j}}=m^2_{3/2} \left[ \delta_{i
{\bar j}} + \frac{n_i}{(T + \overline{T})^2} |G^T|^2\delta_{i{\bar
j}} -|G^+|^2(T + \overline{T})^{m_{ij}+{n_i-n_j \over 2}}
Z'_{i\overline{j}} \right. \nonumber \\
&& \left. -|G^-|^2(T + \overline{T})^{p_{ij}+{n_i-n_j \over 2}}
Z''_{i\overline{j}} \right] \ . \label{smasses}
\end{eqnarray}

As for the gauginos, for the phenomenological analysis we study in
detail the universal case, where the gravity-mediated contributions
are dominated by the term
\begin{equation}
(m_i^{\mathrm{Grav.}})^2 = ({\tilde m}_0^2)_{i {\bar j}} \simeq
m_{3/2}^2 \delta_{i {\bar j}}.
\end{equation}
This is actually a strong assumption, since whereas the contribution
to the scalar masses coming from the moduli and the field
$\Phi_-$ are suppressed compared to the universal first term, we have
not enough information about the third term in the rhs of (\ref{smasses}). Indeed, this term is negligibly small if $r_{ij} \equiv {m_{ij}+{(n_i-n_j) / 2}}
\leq -1$, whereas it is comparable to the universal contribution for
$r_{ij}=0$ and dominant for $r_{ij} > 0$. Whereas this last case
cannot arise in a string compactification, the case  $r_{ij}=0$
could and deserve a more detailed study from the viewpoint of
possible flavor-dependent $\Phi_+$ couplings.

%%%%%%%%%%%%%%%%%%%%%%%%%%%%%%%%%%%%%
%%%%%%%%%%%%%%%%%%%%%%%%%%%%%%%%%%%%%

\subsection{Standard and Non-standard GMSB contributions}

The exact calculation of the radiatively induced gaugino and scalar
masses, due to one messenger multiplet, gives
\cite{gr} for the gaugino mass

\begin{equation}
M^{\mathrm{GMSB}}_a=\frac{g_a^2 m_f S_Q}{8 \pi^2} \ \frac{y_-
\log{y_-}-y_+\log{y_+}-y_- y_+ \log{(y_-/y_+)}}{(y_--1)(y_+-1)}
\end{equation}

\noindent and for the scalar masses \cite{Poppitz:1996xw,gr}

\begin{equation}
(m_i^{\mathrm{GMSB}})^2 \ = \  \frac{C_i~S_Q}{128 \pi^4}m_f^2
 \ F(y_-,y_+,\Lambda_{\mathrm{UV}}^2/m_f^2) \ , 
\end{equation}

\noindent with $y_i=m_i^2/m_f^2$ and $m_i, i=\pm$ the two scalar messengers mass eigenvalues. The function $F$ is given by

\begin{eqnarray}
F(y_-,y_+,\Lambda_{\mathrm{UV}}^2/m_f^2)&=& -(2
y_-+2y_+-4)\log{\frac{\Lambda^2_{\mathrm{UV}}}{m_f^2}} \nonumber
\\
&+& 2(2y_-+2y_+-4) +(y_-+y_+)\log{y_-}\log{y_+} \nonumber
\\
&+& G(y_-,y_+) + G(y_+,y_-) \ ,
\end{eqnarray}

\noindent where

\begin{eqnarray}
G(y_-,y_+)&=&  2y_-\log{y_-}+(1+y_-)
\log^2{y_-}-\frac{1}{2}(y_-+y_+)\log^2{y_-} \nonumber
\\
&+& 2(1-y_-)\mathrm{Li_2}(1-\frac{1}{y_-}) + 2
(1+y_-)\mathrm{Li_2}(1-{y_-}) \nonumber
\\
&-& y_- \mathrm{Li_2}(1-\frac{y_-}{y_+}) \ .
\end{eqnarray}

\noindent $\mathrm{Li_2}(x)$ refers to the dilogarithmic function,
defined by  $\mathrm{Li_2}(x)= - \int_0^1 dz z^{-1}\log{(1-xz)}$. If
\begin{equation}
\lambda F_S , D \ll \lambda^2 \langle S \rangle^2 \ ,
\end{equation}
in (\ref{messmasses}), then it is easy to see that one can expand
the above expressions and have a simplified one in terms of the
ratio $\left| \frac{F_S}{\langle S \rangle}\right|$ and $D$. In
particular, coming back to our model (\ref{superpotential}),  we can
distinguish between the different contributions from the
(positively and negatively) charged messengers and the neutral ones.
In what follows we will distinguish
between the ``standard'' and the ``non-standard'' contributions. We give
the general result in terms of F-terms and vev's, and also the
approximate result in the case $p=0$ \footnote{The case $p > 0$ is
difficult to study in details, since the lack of analytic formula.
Nonetheless, qualitatively it is possible to say that since we
expect in that case $\langle \Phi_+ \rangle \sim \langle \Phi_-
\rangle$ (whereas for $p=0$, $\langle \Phi_+ \rangle \sim 10^{-1}
\langle \Phi_- \rangle$), all the quantities related to the ratio
$\frac{\langle \Phi_- \rangle}{\langle \Phi_+ \rangle}$ will
decrease by a factor 10. In particular we expect that the standard
contributions for the soft masses coming from all the messengers
will be of the same order, and moreover smaller by an order of magnitude 
with respect to the contributions in the case $p=0$.}.
For a coupling $\Phi_-^{n_i'} M_{(+)i'}
\tilde{M}_{(+)i'}$ the charge for both the messengers is denoted by $\left(+
\frac{n_i'}{2} \right)$, whereas for a coupling $\Phi_-^{m_j'}
M_{(-)j'} \tilde{M}_{(-)j'}$ by $\left(- \frac{m_j'}{2}\right)$.

%%%%%%%%%%%%%%%%%%%%%%%%%%%

\paragraph{Messengers with positive charge:}  A messenger $i'$ contributes with
\begin{eqnarray}
M^{\mathrm{GMSB}}_a &=&\frac{g_a^2 S_Q}{8 \pi^2} n_{i'} 
  \frac{F_-}{\langle 
\Phi_- \rangle} \ \simeq \ \frac{g_a^2 S_Q}{8 \pi^2} n_{i'} \
m_{3/2} \ , \\
&& \nonumber \\
&& \nonumber \\
%\end{equation}
%\begin{equation}
(m_i^{\mathrm{GMSB}})^2_{{\mathrm{Stand.}}} &=& \frac{C_i S_Q}{64
  \pi^4} n_{i'}^2 
\left| \frac{F_-}{\langle \Phi_- \rangle}\right|^2 \simeq \frac{C_i
  S_Q}{64 \pi^4} n_{i'}^2 \ m_{3/2}^2 \ , \\
&& \nonumber \\
&& \nonumber \\
%\end{equation}
%\begin{equation}
(m_i^{\mathrm{GMSB}})^2_{{\mathrm{Non-Stand.}}} &=& - \frac{C_i S_Q}{64 \pi^4}  n_{i'} \left( g^2_X D\right)  \left[ \log{\left( \frac{\Lambda_{UV}^2}{(m_f)_{i'}^2} \right)} -1 \right] \nonumber \\
&\simeq& - \frac{C_i S_Q}{32 \pi^4} n_{i'}^2 \frac{m_{3/2}^2}{q}
\left[ \log{\left( \frac{1}{m_{3/2}}\right)}  + \kappa\right]  \left[
  \log{\left( \frac{\Lambda_{UV}^2}{(m_f)_{i'}^2} \right)} -1 \right]
\ , \nonumber \\
\end{eqnarray}
where  the fermionic mass is given by $(m_f)_{i'}^2 \simeq
\lambda^2_{(+)i'} \left( \frac{3 q}{2}\right)^{n_{i'}}\left[
\log{\left( \frac{1}{m_{3/2}}\right)}  + \kappa\right]^{-n_{i'}} $.

%%%%%%%%%%%%%%%%%%%%%%%%%%%

\paragraph{Messengers with negative charge:} A messenger $j'$ contributes with
\begin{eqnarray}
M^{\mathrm{GMSB}}_a&=&\frac{g_a^2 S_Q}{8 \pi^2} m_{j'} 
\frac{F_+}{\langle \Phi_+ \rangle}
\simeq \frac{g_a^2 S_Q}{4 \pi^2} n_{i'} \frac{m_{3/2}}{q} \left[
  \log{\left( \frac{1}{m_{3/2}}\right)}  + \kappa\right]  \\
&& \nonumber \\
&& \nonumber \\
%\end{equation}
%\begin{equation}
(m_i^{\mathrm{GMSB}})^2_{{\mathrm{Stand.}}} &=& \frac{C_i S_Q}{64 \pi^4} m_{j'}^2 \left| \frac{F_+}{\langle \Phi_+ \rangle}\right|^2 \simeq \frac{C_i S_Q}{16 \pi^4} n_{i'}^2 \frac{m_{3/2}^2}{q^2}   \left[ \log{\left( \frac{1}{m_{3/2}}\right)}  + \kappa\right] ^2\\
&& \nonumber \\
&& \nonumber \\
%\end{equation}
%\begin{equation}
(m_i^{\mathrm{GMSB}})^2_{{\mathrm{Non-Stand.}}} &=& \frac{C_i S_Q}{64 \pi^4}m_{j'}  \left( g^2_X D\right) \left[ \log{\left( \frac{\Lambda_{UV}^2}{(m_f)_{j'}^2} \right)} -1 \right] \nonumber  \\
&\simeq& \frac{C_i S_Q}{32 \pi^4} m_{j'}^2 \frac{m_{3/2}^2}{q}
\left[ \log{\left( \frac{1}{m_{3/2}}\right)}  + \kappa\right]  \left[
  \log{\left( \frac{\Lambda_{UV}^2}{(m_f)_{j'}^2} \right)} -1 \right]
\ , \nonumber \\
\end{eqnarray}
where the fermionic mass is given by $(m_f)_{j'}^2 \simeq
\lambda^2_{(-)j'} \frac{1}{3} \left( \frac{3 q}{2}\right)^{2
m_{j'}}\left[ \log{\left( \frac{1}{m_{3/2}}\right)}  +
\kappa\right]^{-2 m_{j'}} $.

%%%%%%%%%%%%%%%%%%%%%%%%%%%

\paragraph{Messengers with zero charge:}  A messenger $r'$ contributes with
\begin{eqnarray}
M^{\mathrm{GMSB}}_a &=&\frac{g_a^2 S_Q}{8 \pi^2} l_{r'} c_{r'} \left(  \frac{F_+}{\langle \Phi_+ \rangle} +\frac{F_-}{\langle \Phi_- \rangle}\right) \nonumber \\
& \simeq& \frac{g_a^2 S_Q}{4 \pi^2} l_{r'} c_{r'} \frac{m_{3/2}}{q}
\left[ \log{\left( \frac{1}{m_{3/2}}\right)}  + \kappa
  +\frac{q}{2}\right] \ , \\
&& \nonumber \\
&& \nonumber \\
(m_i^{\mathrm{GMSB}})^2_{{\mathrm{Stand.}}} &=& \frac{C_i S_Q}{64 \pi^4} l_{r'}^2 c_{r'}^2 \left( \left| \frac{F_+}{\langle \Phi_+ \rangle} +  \frac{F_-}{\langle \Phi_- \rangle}\right| \right)^2 \nonumber \\
& \simeq&  \frac{C_i S_Q}{16 \pi^4} l_{r'}^2 c_{r'}^{2}
\frac{m_{3/2}^2}{q^2} \left[ \log{\left( \frac{1}{m_{3/2}}\right)}
+ \kappa +\frac{q}{2}\right] ^2 \ , 
\end{eqnarray}

where $c_{r'} = \left[1 + \frac{\mu_{r'}}{\lambda_{0r'} \left( \Phi_+ \Phi_- \right)^{l_{r'}}} \right]^{-1}$.\\
In this case there is no non-standard contribution for the scalar
soft masses, since the supertrace for these messengers vanishes.

%%%%%%%%%%%%%%%%%%%%%%%%%%%%%%%%%%%%%%%%%%%%%%%%%%%%%%%%%%%%%%%%%%%%%%

%%%%%%%%%%%%%%%%%%%%%%%%%%%%%%%%%%%%%%%%
\subsection{Anomaly mediation contributions}

The contributions to the terms soft coming from the anomaly
mediation mechanism, are typically of order
\begin{equation}
m^{Anom.}_i \sim \frac{g^2}{16 \pi^2} m_{3/2}
\end{equation}
and then are naturally suppressed in our case already by the
universal terms coming from the gravity mediation mechanism for
$\tilde{\epsilon} \gtrsim{\cal O} \left(10^{-1}
 \right)$.\\
It is interesting to note that this is different from the scenario studied in \cite{Choi:2004sx}
%and from the ``sequestered'' cases, 
where the gravity mediation
contributions are suppressed with respect to the gravitino mass by a
loop factor, and then become of the same
order of those produced by anomaly mediation, allowing the so-called
mirage unification.

\subsection{Explicit soft terms}

We explicit in the appendix the charge dependence of the soft breaking terms
at high scale for the scalar fields, in the framework of our phenomenological parametrization.

\begin{eqnarray}
m_{Q_L}^2&=& m_{3/2}^2 \left\{ 1 +
S_Q \left( - \beta + \frac{\alpha^2}{N} \right)
\left[
\frac{4}{3}  g_3^4
+\frac{3}{4}  g_2^4 + \frac{3}{5}
\left(
\frac{1}{6}
\right)^2
g_1^4
\right] \right\},
\nonumber \\
m_{U_R}^2&=& m_{3/2}^2 \left\{ 1 +
S_Q \left( - \beta + \frac{\alpha^2}{N} \right)
\left[
\frac{4}{3}  g_3^4
+ \frac{3}{5}
\left(
\frac{-2}{3}
\right)^2
g_1^4
\right] \right\},
\nonumber \\
m_{D_R}^2&=& m_{3/2}^2 \left\{ 1 +
S_Q \left( - \beta + \frac{\alpha^2}{N} \right)
\left[
\frac{4}{3}  g_3^4
+ \frac{3}{5}
\left(
\frac{1}{3}
\right)^2
g_1^4
\right] \right\},
\nonumber \\
m_{E_L}^2&=& m_{3/2}^2 \left\{ 1 +
S_Q \left( - \beta + \frac{\alpha^2}{N} \right)
\left[
+\frac{3}{4} g_2^4 + \frac{3}{5}
\left(
\frac{-1}{2}
\right)^2
g_1^4
\right] \right\},
\nonumber \\
m_{E_R}^2&=& m_{3/2}^2 \left\{ 1 +
S_Q \left( - \beta + \frac{\alpha^2}{N} \right)
\left[
\frac{3}{5}
\left(
1
\right)^2
g_1^4
\right] \right\},
\nonumber \\
m_{H_u}^2&=& m_{3/2}^2 \left\{ 1 +
S_Q \left( - \beta + \frac{\alpha^2}{N} \right)
\left[
\frac{3}{4}  g_2^4 + \frac{3}{5}
\left(
\frac{1}{2}
\right)^2
g_1^4
\right] \right\},
\nonumber \\
m_{H_d}^2&=& m_{3/2}^2 \left\{ 1 +
S_Q \left( - \beta + \frac{\alpha^2}{N} \right)
\left[
\frac{3}{4} g_2^4 + \frac{3}{5}
\left(
\frac{-1}{2}
\right)^2
g_1^4
\right] \right\},
\end{eqnarray}

\noindent
where as usual $S_Q$ is the Dynkin index of the messenger representation ($S_Q = 1/2$ for the fundamental representation), $g_i$ are the gauge couplings at GUT scale
($\alpha_i=g_i^2/16\pi^2$)
whereas
$\alpha$, $\beta$ and $N$ have been defined in Sections \ref{micro} and \ref{pheno}. In all our discussion we assumed for simplicity the messengers in complete representations of $SU(5)$.

%%%%%%%%%%%%%%%%%%%%%%%%%%%%%%%%%%%%%%%%%%%%%%%%%%%%%%%%%%%%%%%%%%%%%%%%%%%%%%%%%%%%%%%%%%%%%%%

\nocite{}
\bibliography{bmn}
\bibliographystyle{unsrt}

\end{document}